\documentclass[aps,prc,twocolumn,preprintnumbers,amsmath,amssymb,amsfonts,nofootinbib,superscriptaddress,altaffilletter]{revtex4-2}

\usepackage{graphicx}
\usepackage[unicode]{hyperref}
\hypersetup{
  bookmarksopen=true
}
\usepackage{amssymb}
\usepackage{amsmath}
\usepackage{dsfont}
\usepackage{braket}
\usepackage{mathtools, lipsum}
\usepackage{color}
\usepackage{supertabular}
\usepackage{dcolumn}
\usepackage{multirow}
\usepackage{upgreek}
\usepackage{xcolor}
\usepackage{booktabs}
\usepackage{natbib}
\usepackage{float}
\usepackage{orcidlink}

\newif\ifshowfigs
\showfigstrue

\usepackage{soul}

\newcommand{\useMSU}[1]{}

\begin{document}

\title{A unified mean-field description of dense hadron matter constrained from $\chi$EFT and lattice QCD}

\author{Md. Emanuel Hoque\,\orcidlink{0009-0002-8488-8758}}
\email{emanuel.hoque@saha.ac.in}
\affiliation{Saha Institute of Nuclear Physics, 1/AF Bidhannagar, Kolkata-700064, India}
\affiliation{Homi Bhabha National Institute, Anushakti Nagar, Mumbai 400094, India}

\author{Arunava Mukherjee\,\orcidlink{0000-0003-1274-5846}}
\email{arunava.mukherjee@saha.ac.in}
\affiliation{Saha Institute of Nuclear Physics, 1/AF Bidhannagar, Kolkata-700064, India}
\affiliation{Homi Bhabha National Institute, Anushakti Nagar, Mumbai 400094, India}

\author{Sayantan Sharma\, \orcidlink{0000-0001-5775-4805}}
\email{sayantans@imsc.res.in}
\affiliation{The Institute of Mathematical Sciences, Taramani, Chennai 600113, India}
\affiliation{Homi Bhabha National Institute, Anushakti Nagar, Mumbai 400094, India}

\begin{abstract}
We construct a relativistic mean field theory (RMFT) of hadrons in an attempt to describe the confined phase of QCD across a wide range of temperatures and baryon densities. This includes the scalar and vector mesons interacting with baryons and resonances up to $3$ GeV, including those which are not yet discovered but predicted within the quark model and lattice QCD. The baryon--meson couplings are constrained through a two-stage Bayesian analysis. In the first stage, we constrain six different couplings in our theory using six nuclear empirical parameters inferred from the chiral effective field theory predictions of the energy per nucleon for pure neutron matter over baryon densities $n_B \in [0.08, 0.32]~\mathrm{fm}^{-3}$, employing a Gaussian-process likelihood to account for correlations in the input data. In the second stage, this posterior is set as the prior for constraining different baryon--meson couplings in our RMFT construction, using lattice QCD input. The resulting posterior provides substantially improved constraints for different baryon--meson couplings and non-linear meson self-interactions, which are consistent with the current astrophysical observations of neutron star masses, radii and tidal deformability parameters.
\end{abstract}

\maketitle

\section{Introduction}
\label{sec:introduction}

An \textit{ab initio} determination of the properties of hadrons and their interaction channels from QCD, the fundamental theory of strong interactions, remains challenging~\cite{Lattimer:2015nhk, Sorensen:2023zkk}. The recent detection of gravitational waves by the LIGO/Virgo detector network~\cite{LIGOScientific:2017vwq, GW170817_mma2017, GW190425_ApJ2020} originating from binary neutron star (BNS) merger events GW170817 and GW190425, followed by mass-radius constraints for several X-ray pulsars from NICER observations~\cite{Miller_2019, Riley_2019, Miller_2021, Riley_2021, Dittmann_2024, Salmi_2024, Choudhury_2024}, provided impetus for a more comprehensive understanding of the dense nuclear equation of state~\cite{Annala:2017llu, De_etal_prl2018, GW170817_eos_mr_prl2018, Capano:2019eae, Dietrich_etal_science2020}. Though lattice QCD is the most successful non-perturbative framework that has provided important predictions about the masses and couplings among hadrons~\cite{BMW:2008jgk, Hoelbling:2010syv, Aoki:2012tk}, traditional lattice QCD algorithms suffer from the infamous \emph{sign problem}~\cite{deForcrand:2010j3, Aarts_2016, Schmidt_2017, Nagata_2022} at baryon densities relevant for neutron star cores. On the other hand, perturbative techniques are reliable only in a regime where densities are much larger than the nuclear saturation density~\cite{Kurkela_PRD2010, Gorda_PRL2021, Komoltsev_2022}. 

Traditionally, cold nuclear matter has been studied by solving the many-body Schr\"{o}dinger equation using quantum Monte Carlo methods (see e.g. Ref.~\cite{Carlson:2014vla}). Although these methods can successfully describe many aspects of low-energy nuclear physics, such as observing hints of a liquid-gas phase transition~\cite{Muller:1999cp}, providing a renormalizable quantum theory based on relativistic hadrodynamics~\cite{WALECKA1974491} allows for a closer connection to QCD. Though an \textit{ab initio} Monte Carlo lattice study of a relativistic hadron model has been formulated~\cite{Brockmann:1992in}, it is currently possible to include only a few baryon and meson degrees of freedom. Thus, mean-field analysis and its nonlinear extensions~\cite{WALECKA1974491, BOGUTA1977413, Serot:1984ey}, referred to as relativistic mean-field theory (RMFT), are typically used to study cold and dense nuclear matter~\cite{Glendenning_book2000, Dutra_PRC2014}, most notably in understanding the nuclear liquid-gas transition~\cite{Serot:1992ti} as well as describing neutron-star interiors~\cite{Oertel_RMP2017, Hornick_PRC2018, Hornick_PRC2021erratum}. The RMFT framework can be extended to study nuclear interactions at finite temperatures as well, enabling a description of binary neutron-star mergers~\cite{Oertel_RMP2017, Shen_NPA1998, Hempel_NPA2010, Steiner_ApJ2013}, and quasi-equilibrium transport phenomena~\cite{Alford_PRL2018, Alford_PRC2019, Alford_universe2021}. The extension of RMFTs to study the chiral phase transition and nuclear potential~\cite{Epelbaum:2008ga} requires the inclusion of pions and pion-nucleon interaction terms, which have been studied within the chiral effective theory framework ($\chi_\mathrm{EFT}$) pioneered by Weinberg~\cite{Weinberg:1991um}, using both perturbative~\cite{Bedaque:2002mn} and non-perturbative techniques~\cite{Lee:2004si}.

The primary limitation of the RMFT framework lies in how accurately all possible baryon--meson interactions are taken into account and how the corresponding couplings are constrained. While couplings between the ground state nucleons and mesons can be conventionally constrained by the nuclear empirical parameters, typically estimated at the saturation density \cite{Dutra_PRC2014, Hornick_PRC2018}, those for hyperons, $\Delta$ resonances, and other heavier baryon states are only weakly constrained by terrestrial experiments. Hypernuclear spectroscopy places constraints on the depth of the $\Lambda$ potential in dense nuclear matter, while providing only qualitative constraints on the $\Sigma$ and $\Xi$ potentials~\cite{Millener_PRC1988, Gal_RMP2016}, whereas the couplings of $\Delta$ resonances and heavier states remain largely unconstrained~\cite{Drago_PRC2014}. Instead, these are typically assigned using symmetry arguments such as $\mathrm{SU}(6)$ spin--flavor relations or quark-counting rules~\cite{Schaffner_PRL1993}. These excited baryons lead to the softening of the equation of state at high densities and moderately high temperatures,  exemplified by the hyperon puzzle~\cite{hyperon_puzzle_Lonardoni_PRL2015}. Hence, it is imperative to put in more stringent theoretical constraints on these couplings.

Hadronic interactions are universal; when properly constrained, they will be able to describe nuclear matter over the entire range of temperatures and baryon densities that characterize the chiral-symmetry-broken (confined) phase of QCD.  As emphasized, different theoretical models describe hadronic interactions in different regimes of the QCD phase diagram. Chiral effective field theory provides systematic predictions for the energy per nucleon of pure neutron matter at zero temperature and up to nearly twice the nuclear saturation density, with quantifiable correlated truncation uncertainties \cite{Lynn_PRL2016, Tews_2018, Drischler_PRL2020, Drischler_2021}. On the other hand, at finite temperature $T$ and baryon chemical potentials $\mu_B<3T$,  lattice QCD can consistently predict the hadronic equation of state up to the chiral crossover transition~\cite{Aoki:2006we, Bazavov_PLB2019} as well as correlations and fluctuations of different conserved charges~\cite{Bazavov_PRD2017, Borsanyi_2018, Borsanyi:2022qlh}. At temperatures lower than the chiral crossover transition, these conserved charge susceptibilities can be very well described~\cite{Bazavov_PRL2014, Bollweg_2021, Alba_PRD2017} by a non-interacting gas of hadrons and all possible resonances, many of which are yet to be discovered experimentally but are known to exist through quark models or lattice QCD calculations. In such simple hadron models too, it is important to include repulsive baryon--meson interactions to explain lattice data for various susceptibilities closer to the chiral crossover, even at small baryon densities~\cite{Albright:2014gva,Vovchenko:2016rkn,Biswas:2024xxh}. Importantly, these two models describe the hadron phase in two complementary regimes. Whereas $\chi_\mathrm{EFT}$ constrains the nucleon couplings at zero temperature and high nuclear densities, the variants of the hadron resonance gas model describe the hadron phase at temperatures below the crossover transition and small densities.

In this work, we construct a mean-field description of the hadronic phase of QCD that is applicable over a wide range of temperatures and baryon densities. By using the nuclear empirical parameters (NEPs) derived from the energy per nucleon described by $\chi_\mathrm{EFT}$ for baryon number densities $n_B \in [0.08, 0.32]~\mathrm{fm}^{-3}$~\cite{Tews_2018} in pure neutron matter and lattice QCD data in the two complementary regimes, we provide more stringent constraints on the baryon--meson couplings, especially in the strange sector. This allows us to improve considerably upon our earlier effort~\cite{Aman_PRD2024} and to extend mean-field studies of baryon-number fluctuations~\cite{Fukushima_PRC2015, Kumar:2026gru}, as well as to complement other recent approaches~\cite{Kumar:2024owe,Kumar:2025rxj}. The plan of our paper is as follows: we begin by introducing our hadron model in the following section and describing in detail our procedure for constraining its couplings. We then discuss the cold, neutral, beta-equilibrated matter within our framework to describe neutron stars. We demonstrate the effectiveness of our procedure in constraining baryon--meson couplings, with which mass-radius bounds for neutron stars are found to be consistent with the existing astrophysical data. We then conclude by discussing the implications of our work.

\section{Mean-field description of relativistic hadrodynamics}
\label{sec:model}

We start by describing the Lagrangian density of interacting hadrons that we will be studying in our work. This is motivated by the original model of hadrodynamics developed by Walecka~\cite{WALECKA1974491}, which describes the nuclear matter near the liquid-gas transition. There are no pions and kaons included in this model, which is acceptable as long as we are not very close to the chiral transition. However, we have made two important extensions of the original Walecka model. Not only have ground-state strange baryons like $\Lambda$ been included as motivated in Refs.~\cite{BuntaGmuca_PRC2004, Alford_PRC2022, Passarella_PRC2025}, but also all other possible strange and non-strange baryons and resonances up to $3$ GeV~\cite{Aman_PRD2024}, which were not included earlier in the original construction. We sometimes refer to these additional baryon states as the tower of states (TOS). This accounts for the inclusion of hadron interactions with resonant attractive channels~\cite{Venugopalan:1992hy}, e.g., pion-nucleon scattering. The baryons interact through the exchange of scalar $\sigma$, vector $\omega^\mu$, and the isovector-vector $\vec{\rho}^\mu$ mesons. In order to account for the hyperon--hyperon attraction observed in $\Lambda\Lambda$ hypernuclei, the hidden strange scalar $\sigma^*$ and vector $\phi^\mu$ mesons are also introduced. The Lagrangian density is compactly summarized as 

\begin{align}
\mathcal{L}={}&
\sum_b\bar{\psi}_b
\Bigg[
i\gamma^\mu\partial_\mu
-m_b
+g_{\sigma b}\sigma
-g_{\omega b}\gamma^\mu\omega_\mu
\nonumber\\
&
\qquad
-g_{\rho b}\gamma^\mu \vec{I}_b\!\cdot\!\vec{\rho}_\mu
+g_{\sigma^{*}b}\sigma^{*}
-g_{\phi b}\gamma^\mu\phi_\mu
\Bigg] ~\psi_b
\nonumber\\
&
+\mathcal{L}_\sigma
+\mathcal{L}_\omega
+\mathcal{L}_\rho
+\mathcal{L}_{\omega\rho}
+\mathcal{L}_{\sigma^{*}}
+\mathcal{L}_\phi~.
\label{eq:lagrangian}
\end{align}

where $b$ denotes all baryons and resonances up to $3$ GeV included from the quark-model-inspired HRG described in Ref.~\cite{Bollweg_2021}. Here, $g_{mb}$ denotes the coupling of the mesons $m\in \{\sigma, \omega, \rho, \sigma^{*}, \phi\}$ to the baryon species $b$. The different ground state mesons and their dynamics are described by the following terms: 

\begin{align}
\mathcal{L}_\sigma &=
\frac12
\Big(
\partial_\mu\sigma\,\partial^\mu\sigma
-m_\sigma^2\sigma^2
\Big)
-\frac{ b_\sigma}{3} m_N(g_\sigma\sigma)^3 -\frac{ c_\sigma}{4}(g_\sigma\sigma)^4,
\\[0.3em]
\mathcal{L}_\omega &=
-\frac14\omega_{\mu\nu}\omega^{\mu\nu}
+\frac12 m_\omega^2~\omega_\mu\omega^\mu,
\\[0.3em]
\mathcal{L}_\rho &=
-\frac14\vec{\rho}_{\mu\nu}\!\cdot\!\vec{\rho}^{\mu\nu}
+\frac12 m_\rho^2~\vec{\rho}_\mu\!\cdot\!\vec{\rho}^\mu,
\\[0.3em]
\mathcal{L}_{\omega\rho} &=
\frac12\Lambda_V
\Big(g_\omega^2~\omega_\mu\omega^\mu\Big)
\Big(g_\rho^2~\vec{\rho}_\mu \cdot \vec{\rho}^\mu\Big),
\\[0.3em]
\mathcal{L}_{\sigma^{*}} &=
\frac12
\Big(
\partial_\mu\sigma^{*}\partial^\mu\sigma^{*}
-m_{\sigma^{*}}^2~\sigma^{*2}
\Big),
\\[0.3em]
\mathcal{L}_\phi &=
-\frac14\phi_{\mu\nu}\phi^{\mu\nu}
+\frac12 m_\phi^2~\phi_\mu\phi^\mu .
\end{align}

in terms of the antisymmetric field tensors for the vector fields,

\begin{align}
\omega_{\mu\nu} &= \partial_\mu \omega_\nu - \partial_\nu \omega_\mu, ~\vec{\rho}_{\mu\nu} = \partial_\mu \vec{\rho}_\nu - \partial_\nu \vec{\rho}_\mu, \\
\phi_{\mu\nu} &= \partial_\mu \phi_\nu - \partial_\nu \phi_\mu.
\end{align}

The couplings without a baryon index appearing in the meson self-interaction terms are those of the corresponding meson with the nucleons only, i.e., $g_\sigma \equiv g_{\sigma N}$, $g_\omega \equiv g_{\omega N}$, and $g_\rho \equiv g_{\rho N}$. The scalar meson $\sigma$ mediates an intermediate-range attractive interaction that lowers the effective mass of the baryons within the mean-field approximation, while the vector meson $\omega$ mediates a short-range repulsion that leads to the stiffening of the nuclear equation of state (EOS) at high density. We only consider the coupling of the isovector $\rho$ to the third isospin component of each baryon ($I_{3} = \pm\tfrac12$ for nucleons), which in turn quantifies the energy cost of having a finite isospin asymmetry. The cubic and quartic scalar self-interactions of the $\sigma$ meson field, with coupling strengths $b_\sigma$ and $c_\sigma$ respectively, soften the scalar potential~\cite{BOGUTA1977413} and are required to reproduce the incompressibility and the effective mass of the nucleons at the saturation density. The $\omega$--$\rho$ coupling strength $\Lambda_V$~\cite{Horowitz_PRL2001} modifies the density dependence of the symmetry energy, primarily its slope, which is denoted by $L_{\rm sym}$. The three-body hyperon-nucleon interaction is believed to play a fundamental role in the softening of the equation of state and in the consequent reduction of the predicted maximum mass of a neutron star~\cite{hyperon_puzzle_Lonardoni_PRL2015}. Hidden-strange mesons $\sigma^{*}$ and $\phi_\mu$~\cite{Schaffner_PRL1993} couple only to strange baryons, providing additional attractive and repulsive interactions, respectively, among hyperons; thus $g_{\sigma^{*}b} = g_{\phi b} = 0$ when $b$ denotes nucleons and non-strange resonances. The three non-strange meson masses are fixed according to the values used in the IU-FSU RMFT parameter set, i.e., $m_{\sigma} = 491.5~\text{MeV}$, $m_{\omega} = 782.5~\text{MeV}$, and $m_{\rho} = 763.0~\text{MeV}$ \cite{Fattoyev_PRC2010, Alford_PRC2022}. The strange meson masses are set to $m_{\sigma^{*}} = 975~\text{MeV}$ and $m_{\phi} = 1020~\text{MeV}$~\cite{Schaffner_PRC1996, Zhao_PRC2015}, respectively, and the vacuum nucleon mass is set to $m_N = 939~\text{MeV}$.

For uniform static nuclear matter in the mean-field approximation, only the following thermal expectation values of the meson  fields are considered,  $\bar{\sigma}$, $\overline{\sigma^{*}}$, $\overline{\omega_0}$, $\overline{\rho_{03}}$, and $\overline{\phi_0}$. In this approximation, the quantum fluctuations are neglected. Each baryon species then acquires an effective mass and an effective chemical potential denoted as

\begin{align}
m^{*}_b &= m_b - g_{\sigma b}~\bar{\sigma} - g_{\sigma^{*}b}~\overline{\sigma^{*}},
\label{eq:effmass}\\
\mu^{*}_b &= \mu_b - g_{\omega b}~\overline{\omega_0}
             - g_{\rho b}\,I_{3b}\,\overline{\rho_{03}}
             - g_{\phi b}~\overline{\phi_0} ,
\label{eq:effmu}
\end{align}

Since each baryon $b$ carries baryon number $B_b$, charge $Q_b$ and strangeness $S_b$, the chemical potential becomes $\mu_b = B_b \mu_B + Q_b\mu_Q + S_b\mu_S$. The classical 
equations of motion of the meson fields can be derived from Eq.~\eqref{eq:lagrangian},

\begin{align}
m_\sigma^2\bar{\sigma}
 + b_\sigma m_N g_\sigma^3 \bar{\sigma}^2
 + c_\sigma g_\sigma^4 \bar{\sigma}^3
 &= \sum_b g_{\sigma b}\, n^{s}_b ,
\label{eq:eom_sigma}\\
m_{\sigma^{*}}^2\overline{\sigma^{*}}
 &= \sum_b g_{\sigma^{*} b}\, n^{s}_b ,
\label{eq:eom_sigmastar}\\
m_\omega^2~\overline{\omega_0}
 + \Lambda_V ~g_\omega^2~ g_\rho^2\, \overline{\rho_{03}}^2\, \overline{\omega_0}
 &= \sum_b g_{\omega b}\, n_b ,
\label{eq:eom_omega}\\
m_\rho^2\overline{\rho_{03}}
 + \Lambda_V g_\omega^2 g_\rho^2\, \overline{\omega_0}^2\, \overline{\rho_{03}}
 &= \sum_b g_{\rho b}\,I_{3b}\, n_b ,
\label{eq:eom_rho}\\
m_\phi^2\overline{\phi_0} &= \sum_b g_{\phi b}\, n_b .
\label{eq:eom_phi}
\end{align}

where the source terms are expressed in terms of the scalar and vector number densities at temperature $T$, defined as:

\begin{align}
n_b &= \frac{\gamma_b}{2\pi^2}\int_0^{\infty}\! dk\, k^2
        \Big[ f^{+}_b(k) - f^{-}_b(k) \Big],
\label{eq:ndens}\\
n^{s}_b &= \frac{\gamma_b}{2\pi^2}\int_0^{\infty}\! dk\, k^2\,
        \frac{m^{*}_b}{E^{*}_b(k)}
        \Big[ f^{+}_b(k) + f^{-}_b(k) \Big],
\label{eq:sdens}
\end{align}

Here, $\gamma_b = 2J_b+1$ denotes the spin degeneracy of baryon species $b$, which carries an effective energy $E^{*}_b(k)=\sqrt{k^2+m^{*2}_b}$, and $f^{\pm}_b(k) =1/\left( 1 + \mathrm{e}^{\frac{E^{*}_b(k) \mp \mu^{*}_b}{T}}\right)$ represent the particle and antiparticle occupation numbers, respectively. Finally, the pressure is obtained in terms of the mean fields as follows,

\begin{align}
p={}& \sum_b \frac{\gamma_b}{6\pi^2}\int_0^{\infty}\! dk\,
        \frac{k^4}{E^{*}_b(k)}
        \left[ f^{+}_b(k) + f^{-}_b(k) \right]
\nonumber\\
&-\frac12 m_\sigma^2\bar{\sigma}^2
 -\frac13 b_\sigma m_N (g_\sigma\bar{\sigma})^3
 -\frac14 c_\sigma (g_\sigma\bar{\sigma})^4
 -\frac12 m_{\sigma^{*}}^2~\overline{\sigma^*}^{2}
\nonumber\\
&+\frac12 m_\omega^2~\overline{\omega_0}^2
 +\frac12 m_\rho^2~\overline{\rho_{03}}^2
 +\frac12 m_\phi^2~\overline{\phi_0}^2
\nonumber\\
&+\frac12 \Lambda_V~ g_\omega^2 ~g_\rho^2~\,
   \overline{\omega_0}^2\,\overline{\rho_{03}}^2.
\label{eq:pressure}
\end{align}

The energy density $\varepsilon$ can now be obtained from the Euler relation, 
$\varepsilon = -p + Ts + \sum_b \mu_b n_b$.

\section{Methodology}
\label{sec:constraints}

\subsection{Constraints on couplings from $\chi_\mathrm{EFT}$}
\label{subsec:constraints_chiEFT}

We first constrain the ground state nucleon--meson couplings within our RMFT by requiring that it reproduces the low-density behavior of pure neutron matter within the $\chi_\mathrm{EFT}$. We choose a set of six nuclear empirical parameters (NEPs) estimated at the nuclear saturation density,

\begin{equation}
\mathbf{\theta}_{\rm NEP} \equiv
\{\, E_{\rm sat},\ n_{\rm sat},\ \frac{m^*}{m},\ K_{\rm sat},\ E_{\rm sym},\ L_{\rm sym}\,\},
\label{eq:neps}
\end{equation}

which are in one-to-one correspondence with the nucleon and self-interaction couplings $\{g_{\sigma N}, g_{\omega N}, g_{\rho N}, b_\sigma, c_\sigma, \Lambda_V\}$. The NEP denoted by $K_{\rm sat}$ measures the curvature of the energy per nucleon $E/A$ of nuclear matter as a function of the nucleon density, $n_B$, about the saturation density $n_{\rm sat}$, 

\begin{equation}
E_{\rm sat}
=
\left.\frac{E}{A}\right|_{n_{\rm sat}} - m_N,
\label{eq:Esat}
\end{equation}

\begin{equation}
K_{\rm sat}
=
9n_{\rm sat}^2
\left.
\frac{\partial^2(E/A)}{\partial n_B^2}
\right|_{n_{\rm sat}},
\label{eq:Ksat}
\end{equation}

Apart from the mass ratio, the two remaining NEPs are the symmetry energy and its slope at the saturation density,
\begin{equation}
\frac{m^*}{m} = \frac{m_N - g_{\sigma N}\bar{\sigma}}{m_N}, ~E_{\rm sym}
=
S(n_{\rm sat}),
\label{eq:Esym}
\end{equation}

\begin{equation}
L_{\rm sym}
=
3n_{\rm sat}
\left.
\frac{\partial S}{\partial n_B}
\right|_{n_{\rm sat}}~.
\label{eq:Lsym}
\end{equation}

\noindent
Here, $S(n_B)$ is the density-dependent symmetry energy of the nuclear matter and is the difference between the energy per nucleon of pure neutron matter and symmetric nuclear matter at a given baryon number density $n_B$. 

The priors for six NEPs are presented in Table~\ref{tab:nep_priors}. $E_{\rm sat}$ and $n_{\rm sat}$ are adapted from Refs.~\cite{Margueron_PRC2018, Passarella_PRC2025}; $K_{\rm sat}$ is chosen to enclose $240 \pm 20\text{ MeV}$~\cite{Garg_2018}; and the ranges for the isovector quantities are kept agnostic~\cite{Lattimer_2013, Lattimer_2014, Drischler_PRL2020} to allow $\chi_\mathrm{EFT}$ estimates to determine their posteriors. The effective Dirac mass spans the range of different RMFT parameterizations~\cite{Dutra_PRC2014, Hornick_PRC2018, Hornick_PRC2021erratum}. Following~\citet{Li_2025}---which notes that a larger effective mass implies the scalar and vector potentials are too weak to reproduce the empirical saturation point of symmetric nuclear matter---the upper limit is set at $m^*/m = 0.80$. This corresponds to one of the higher values attainable in RMFTs. All priors are sampled uniformly over the interval defined by the lower and upper bounds of each parameter.

\begin{table*}[]
\centering
\caption{The priors for NEPs are sampled uniformly over a range set by the lower and upper limits for each corresponding parameter summarized below.}
\label{tab:nep_priors}
\resizebox{0.65\textwidth}{!}{%
\begin{tabular}{|c|cccccc|}
\toprule
Parameters & $E_{\rm sat}$ [MeV] & $n_{\rm sat}$ [fm$^{-3}$] & $m^*/m$ & $K_{\rm sat}$ [MeV] & $E_{\rm sym}$ [MeV] & $L_{\rm sym}$ [MeV] \\
\hline
\hline
Lower Limit & $-16.4$ & $0.145$ & $0.55$ & $180$ & $24$ & $0$ \\
\hline
Upper Limit & $-15.2$ & $0.165$ & $0.80$ & $320$ & $36$ & $100$ \\
\bottomrule
\end{tabular}%
}
\end{table*}

\begin{figure}[htbp]
    \centering
    \includegraphics[width=\columnwidth]{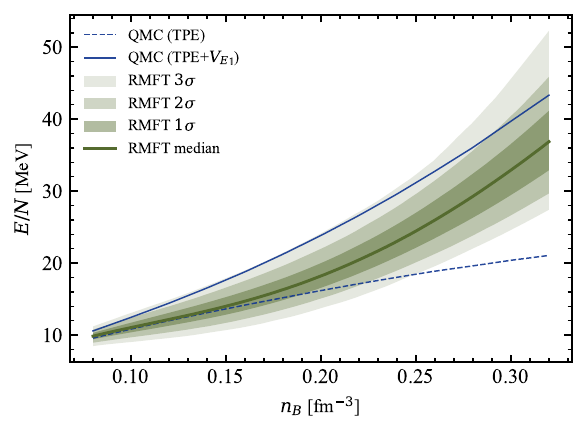}
    \caption{Energy per neutron $E/N$ of pure neutron matter as a function of baryon number density, computed from RMFT with its couplings constrained using $\chi_\mathrm{EFT}$ data. Dark-to-light green shading denotes the $1\sigma$, $2\sigma$, and $3\sigma$ regions, respectively, for the RMFT prediction, along with the median values indicated by the solid green line. The dotted and solid blue curves denote the constraints at NNLO for pure neutron matter obtained within $\chi_\mathrm{EFT}$ with two-pion-exchange (TPE) interaction alone and with the TPE supplemented by the $V_{E\mathds{1}}$ contact term, respectively. }
    \label{fig:EbyA_pnm}
\end{figure}

The reference for the pure nucleon sector is the $\chi_\mathrm{EFT}$ prediction for the energy per neutron $E/N$, over the density range $n_B \in [0.08, 0.32]~\mathrm{fm}^{-3}$, spanning roughly from half to twice the nuclear saturation density, where there is sufficient theoretical control. The dominant uncertainty in this range arises from the choice of couplings in the nucleon interactions. We bracket the uncertainty from two different calculations, one based on the two-pion-exchange (TPE) interaction and another that additionally includes the three-nucleon contact term $V_{E\mathds{1}}$ at NNLO~\cite{Tews_2018, Alford_PRC2022}. We take the lower bound of the TPE result together with the upper bound of the $\text{TPE}{}+V_{E\mathds{1}}$ result to define a band in the $E/N$--density plane (see Figure~\ref{fig:EbyA_pnm}) and identify the band center,
\begin{equation}
\Bigg\langle{\frac{E}{N}}\Bigg\rangle(n_B) = \tfrac{1}{2}\Bigg[\frac{E}{N}\bigg|^{\rm TPE}_{\rm low}+ \frac{E}{N}\bigg|^{\rm TPE{}+V_{E\mathds{1}}}_{\rm high}\Bigg](n_B) ,
\end{equation}

as the central value, with its half-width
\begin{equation}
\sigma_{\frac{E}{N}}(n_B) = \tfrac{1}{2}\Bigg[\frac{E}{N}\bigg|^{\rm TPE{}+V_{E\mathds{1}}}_{\rm high} - \frac{E}{N}\bigg|^{\rm TPE}_{\rm low}\Bigg](n_B)
\label{eq:band_sigma}
\end{equation}
acting as the $1\sigma$ uncertainty of the constraint at each density. This uncertainty is purely systematic rather than statistical, reflecting truncation errors in the chiral expansion. Therefore, such a measure of error is strongly correlated across the entire range of baryon densities and should result in an error band that is expected to vary smoothly rather than fluctuate independently from one density to the next~\cite{SemposkiDrischler_etal_prc2025, Lee_etal_arXiv260809678}.

To take into account this correlation, we treat the $\chi_\mathrm{EFT}$ constraint as a Gaussian process in the energy per neutron values, with a mean given by the band center $\Big\langle{\frac{E}{N}}\Big\rangle(n_B)$ and a covariance matrix constructed out of the band half-widths $\sigma_{\frac{E}{N}}(n_{Bi})$ and a squared-exponential kernel,
\begin{equation}
C_{ij} = \sigma_{\frac{E}{N}}(n_{Bi})\,\sigma_{\frac{E}{N}}(n_{Bj})\,
         \exp\!\left[-\frac{(n_{Bi} - n_{Bj})^2}{2\,\ell^2}\right],
\label{eq:gp_cov}
\end{equation}
where the diagonal reproduces the pointwise $1\sigma$ band defined in Eq.~\eqref{eq:band_sigma}. The off-diagonal terms correlate neighboring densities over a correlation length $\ell = 0.06~\mathrm{fm}^{-3}$, roughly taken as a quarter of the density range over which the $E/N$ band for the $\chi_\mathrm{EFT}$ of pure neutron matter is quoted. For a given vector comprising the parameter values, the model energy per neutron $\frac{E}{N}\big|_{\rm RMFT}(n_B,\bm{\theta}_{\rm NEP})$ is evaluated on a grid of nuclear density values and scored against this Gaussian process. Up to a constant, the log-likelihood is given by
\begin{equation}
\begin{split}
\ln\mathcal{L}(\bm{\theta}_{\rm NEP})
&= -\tfrac{1}{2}\,\bm{r}^{\mathsf T}\,C^{-1}\,\bm{r},
\\[0.5em]
r_i &= \frac{E}{N}\bigg|_{\rm RMFT}(n_{Bi}) - \Bigg\langle{\frac{E}{N}}\Bigg\rangle(n_{Bi}).
\end{split}
\label{eq:gp_like}
\end{equation}

We show the comparison between the resulting posterior-predictive energy per neutron and the $\chi_\mathrm{EFT}$ constraints at NNLO which lies within the blue curves in Figure~\ref{fig:EbyA_pnm}. The dotted blue curve denotes $E/N$ of neutron-matter calculated within $\chi_\mathrm{EFT}$ considering TPE interactions, whereas the solid blue line is the energy per neutron when the TPE is supplemented by the $V_{E\mathds{1}}$ contact term. The posterior-predictive energy per neutron follows the $\chi_\mathrm{EFT}$ band, tracking the band smoothly over the density range $n_B\in[0.08,0.32]~\mathrm{fm}^{-3}$, giving confidence in our analysis procedure and that the constrained couplings are consistent with the low-density pure neutron-matter input. The posterior distributions of the six-dimensional NEPs constrained by the $\chi_\mathrm{EFT}$ for pure nucleon matter are shown in Figure~\ref{fig:corner_chieft_6d} in the Appendix. We can observe that the neutron matter properties primarily constrain the isovector sector $E_{\rm sym}=30.3^{+1.4}_{-1.3}~\mathrm{MeV}$, $L_{\rm sym}=33.6^{+6.9}_{-6.1}~\mathrm{MeV}$ to a very good accuracy and the effective mass $m^*/m=0.77^{+0.02}_{-0.04}$ is very close to the upper bound of the prior support. The saturation parameters remain close to the priors, with only a mild preference for larger values of $n_{\rm sat}$ on the chosen prior range.

\subsection{Constraining RMFT parameters additionally using inputs from Lattice QCD}
\label{subsec:constraints_lQCD}

The $\chi_\mathrm{EFT}$ constraints on the energy per neutron allowed us to constrain the ground state baryon--meson couplings of our RMFT. We now would like to constrain the couplings of the higher excited states of these baryons and resonances with the ground state mesons.  For this purpose we use two specific ratios $\chi_4^B/\chi_2^B$ and $\chi_{31}^{BS}/\chi_{11}^{BS}$ obtained from lattice QCD, the former containing information about all baryons and baryonic resonances and the latter only sensitive to the strange baryons. These ratios are calculated deep in the hadron phase described by a temperature $T=135~\mathrm{MeV}$, normalized baryon chemical potential $\hat{\mu}_B=\mu_B/T\in [2.0, 3.0]$ and charge and strangeness chemical potentials set to zero i.e., $\mu_Q=\mu_S=0$. We recall that the (normalized) susceptibilities of different conserved charges are defined as derivatives of the pressure with respect to the corresponding chemical potentials,

\begin{equation}
\label{eq:susceptibilities}
\chi^{BQS}_{ijk} = \frac{\partial^{\,i+j+k}\,(p/T^4)}
{\partial\hat{\mu}_B^{\,i}\,\partial\hat{\mu}_Q^{\,j}\,\partial\hat{\mu}_S^{\,k}}\,,
\qquad \hat{\mu}_{B,Q,S}\equiv \mu_{B,Q,S}/T\,.
\end{equation}

In order to circumvent the \emph{sign problem}, conserved-charge cumulants are typically obtained at vanishing baryon chemical potential~\cite{deForcrand:2010j3, Schmidt_2017} on the lattice and then calculated at finite $\hat{\mu}_B$ by performing a Taylor expansion about $\hat{\mu}_B=0$. The two ratios at finite densities are calculated using up to sixth order susceptibilities, which are known to a fairly good accuracy in the continuum limit, denoted as
\begin{equation}
\label{eq:chi4B_by_chi2B}
\frac{\chi_4^B}{\chi_2^B}(\hat{\mu}_B)
= \frac{\chi_4^B(0) + \tfrac{1}{2}\,\chi_6^B(0)\,\hat{\mu}_B^2 + \mathcal{O}(\hat{\mu}_B^{4})}
       {\chi_2^B(0) + \tfrac{1}{2}\,\chi_4^B(0)\,\hat{\mu}_B^2
        + \tfrac{1}{24}\,\chi_6^B(0)\,\hat{\mu}_B^4 + \mathcal{O}(\hat{\mu}_B^{6})}\,,
\end{equation}

\begin{equation}
\label{eq:chi31BS_by_chi11BS}
\begin{split}
&\frac{\chi_{31}^{BS}}{\chi_{11}^{BS}}(\hat{\mu}_B) \\
&\quad= \frac{\chi_{31}^{BS}(0) + \tfrac{1}{2}\,\chi_{51}^{BS}(0)\,\hat{\mu}_B^2 + \mathcal{O}(\hat{\mu}_B^{4})}
       {\chi_{11}^{BS}(0) + \tfrac{1}{2}\,\chi_{31}^{BS}(0)\,\hat{\mu}_B^2 + \tfrac{1}{24}\,\chi_{51}^{BS}(0)\,\hat{\mu}_B^4 + \mathcal{O}(\hat{\mu}_B^{6})}\,.
\end{split}
\end{equation}

Another motivation for considering ratios instead of individual conserved charge susceptibilities is that finite volume corrections largely cancel out in the former. We retain only those Taylor coefficients that the lattice resolves above their uncertainty. Further details of the systematics involved in the truncation of Taylor series and choice of the $\hat{\mu}_B$ window are mentioned in Appendices~\ref{app:truncation},~\ref{app:window}, respectively. 

We now focus on constraining the couplings between mesons and the excited states of baryons and baryonic resonances. Rather than fixing the additional baryon--meson couplings by SU(6) symmetry, we treat them as free scaling factors relative to the ground state nucleon couplings and constrain them with the lattice ratios described earlier. In order to ensure that the strange and non-strange baryons couple differently to scalar (vector) mesons giving rise to the requisite attractive (repulsive) interactions, we introduce a set of dimensionless scaling parameters $\alpha_\mathrm{s,ns}^{\rm (at)}~ (\alpha_\mathrm{s,ns}^{\rm (r)})$. For non-strange mesons, $m \in \{\sigma, \omega, \rho \}$ coupling to non-strange baryons, which do not belong to the nucleon family, labeled by $R$, the couplings are fixed to
\begin{equation}
\label{eq:tos_nonstrange_baryon}
g_{m R} = \alpha^{\rm (at)}\, g_{m N}~,~~~
g_{m R} = \alpha^{\rm (r)}\, g_{m N}
\end{equation}
in terms of the ground state meson--nucleon coupling $g_{mN}$ whereas the coupling of non-strange mesons to strange baryons and resonances, labeled by $Y$ are denoted as
\begin{equation}
\label{eq:tos_strange_baryon}
g_{m Y} = \alpha^{\,\rm (at)}_{\rm ns}\, g_{m N}, ~~~
g_{m Y} = \alpha^{\,\rm (r)}_{\rm ns}\, g_{m N}.
\end{equation}
The remaining couplings of the hidden-strange mesons $\sigma^{*},\phi$ to all other strange baryons except $\Lambda$-hyperons, denoted as $Y^{'}$, are expressed in terms of the $\Lambda$-hyperon coupling as 
\begin{equation}
\begin{split}
\label{eq:tos_strange_meson}
g_{m Y^{'}} = \alpha^{\,\rm (at)}_{\rm s}\, g_{m \Lambda}, ~~
g_{m Y^{'}} = \alpha^{\,\rm (r)}_{\rm s}\, g_{m \Lambda}~.
\end{split}
\end{equation}
The $g_{m\Lambda}$ is the reference coupling of the hidden-strangeness mesons to the $\Lambda$-hyperon, fixed through the ratios $g_{\sigma^*\Lambda}/g_{\sigma N}$ and $g_{\phi\Lambda}/g_{\omega N}$. These two ratios along with the six scaling coefficients constitute the eight parameters $\Big\{g_{\sigma^*\Lambda}/g_{\sigma N}$, $g_{\phi\Lambda}/g_{\omega N},\alpha^{(\rm at)},\alpha^{(\rm r)},\alpha^{(\rm at)}_{\rm ns},\alpha^{(\rm r)}_{\rm ns},\alpha^{(\rm at)}_{\rm s},\alpha^{(\rm r)}_{\rm s}\Big\}$ in our extended RMFT which will be now constrained by the two lattice QCD ratios.

The six nuclear empirical parameters that are already constrained in the earlier section provide a posterior that serves as the prior for the nucleon couplings. The nuclear empirical parameters are fed as priors through a Gaussian copula of the $\chi_\mathrm{EFT}$ posterior (truncated strictly on the NEP prior boundaries defined in Table~\ref{tab:nep_priors}), while the eight new parameters carry uniform priors over the 8-dimensional unit hypercube $[0,1]^{8}$.

For each parameter vector $\theta$, the model prediction $\mathbf{m}(\theta) = \{r_a(\hat{\mu}_B^{(n)};\theta)\}$ comprises the susceptibility ratios $r_a \in \{\chi_4^B/\chi_2^B,\ \chi_{31}^{BS}/\chi_{11}^{BS}\}$ evaluated at $N$ different values of $\hat{\mu}_B^{(n)}$ spanning the fit window $\hat{\mu}_B \in [2.0,3.0]$. The ratios of susceptibilities calculated within our RMFT are obtained at each $\hat{\mu}_B^{(n)}$  by performing a numerical differentiation of the model pressure with respect to the reduced chemical potentials using Eq.~(\ref{eq:susceptibilities}). The target data $\mathbf{d}$ and its covariance $\mathsf{C}$ are constructed by a Monte Carlo propagation of the cumulant ratios at $\hat{\mu}_B=0$, calculated on the lattice in  Ref.~\cite{Borsanyi_2018}. This is done by performing a sampling of the cumulant values from a Gaussian distribution whose mean is the lattice data close to continuum and whose width is given by the error quoted. The data from every such draw is then extrapolated at finite density using the Eqs.~(\ref{eq:chi4B_by_chi2B}) and (\ref{eq:chi31BS_by_chi11BS}). The components of $\mathbf{d}$ are the median values of the resulting ratio distributions, and $\mathsf{C}$ is the sample covariance.

\begin{figure}[htbp]
\centering
\includegraphics[width=0.95\columnwidth,height=0.65\textwidth]{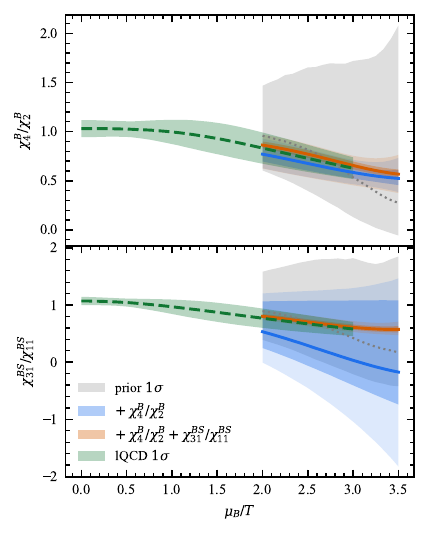}
\caption{
The susceptibility ratios $\chi_4^B/\chi_2^B$ and $\chi_{31}^{BS}/\chi_{11}^{BS}$ are shown as a function of $\mu_B/T$ at $T = 135$~MeV. Each panel shows the model $1\sigma$ prior-predictive band (grey), the posterior-predictive bands with $1\sigma$ and $2\sigma$ credible intervals obtained using $\chi_4^B/\chi_2^B$ (blue) and $\chi_4^B/\chi_2^B+\chi_{31}^{BS}/\chi_{11}^{BS}$ (orange) as inputs, compared to the lattice-QCD $1\sigma$ band (green) obtained using Taylor-series expansion (as discussed in Section~\ref{subsec:constraints_lQCD}).
}
\label{fig:susc_main}
\end{figure}

Since the ratios are reconstructed at different values of $\hat{\mu}_B^{(n)}$, all from the same small set of cumulants at zero density, they are strongly correlated across the fit window. We therefore adopt the full correlated Gaussian likelihood,
\begin{equation}
\label{eq:likelihood}
\begin{aligned}
\ln \mathcal{L}(\theta) &= -\tfrac{1}{2}\,
[\mathbf{m}(\theta) - \mathbf{d}]^{\mathsf{T}}\,
\tilde{\mathsf{C}}^{-1}\,
[\mathbf{m}(\theta) - \mathbf{d}]\,,\\
\tilde{\mathsf{C}} &= \mathsf{C} + \mathrm{diag}\big[(\kappa\, d_n)^2\big]\,.
\end{aligned}
\end{equation}
where the term $\mathrm{diag}\big[(\kappa\, d_n)^2\big]$ is a model-discrepancy floor added in quadrature to the diagonal, and $\kappa = 0.02$ denotes an uncorrelated $2\%$ uncertainty on the predicted ratio at each point. This value of $\kappa$ is chosen to be of the same magnitude as the statistical precision to which the leading cumulants $\chi_2^B$ and $\chi_{11}^{BS}$ are calculated in Ref.~\cite{Borsanyi_2018}. This term primarily regularizes the inversion of the nearly singular covariance matrix $\mathsf{C}$, since the complete uncertainty band is spanned from only a handful of Taylor coefficients at $\hat{\mu}_B = 0$. Additionally, it allows for incorporating numerical errors in the RMFT susceptibilities, lattice estimates and for residual model discrepancy, without diluting the lattice constraint. Posteriors are sampled with dynamic nested sampling as implemented in \texttt{dynesty}~\cite{DYNESTY}. Nested sampling evolves a set of live points drawn from the prior through nested shells of increasing likelihood, and yields the Bayesian evidence $\mathcal{Z}$ together with importance-weighted posterior samples.

Nested sampling requires the prior to be specified as a unit hypercube $[0,1]^{d}$ rather than as a density. This precludes the direct use of a six-dimensional kernel-density estimate of the $\chi_\mathrm{EFT}$-constrained NEP posterior. We therefore impose the $\chi_\mathrm{EFT}$-constrained posterior through a Gaussian copula, which separates the joint distribution into its one-dimensional marginals and a dependence structure. A point $\bm{u}$ of the unit hypercube is mapped to correlated standard normals $\bm{z} = \mathsf{L}\,\Phi^{-1}(\bm{u})$, where $\Phi$ is the standard normal cumulative distribution and $\mathsf{L}$ is the Cholesky factor of the rank correlation matrix of the $\chi_\mathrm{EFT}$ samples. Each component is then mapped onto its NEP according to $\theta_k = F_k^{-1}\!\big(\Phi(z_k)\big)\,,$ where $F_k^{-1}$ is the empirical quantile function of the $k^\mathrm{th}$ NEP. This reproduces the $\chi_\mathrm{EFT}$ marginals exactly and their correlations to a good accuracy. For example, the Pearson coefficient for the quantity $E_{\rm sym}$--$L_{\rm sym}$ is $+0.791$ in the $\chi_\mathrm{EFT}$ samples which agrees well with its value $+0.786$ in the copula draws. The copula also confines the priors within the bounds mentioned in Table~\ref{tab:nep_priors}. The eight new couplings are then mapped linearly onto $[0,1]^{8}$.

We compile our predictions for the two susceptibility ratios across the fit window in $\mu_B/T$, along with the band denoting lattice QCD results which are reconstructed from the truncated Taylor series in Figure~\ref{fig:susc_main}. The gray band denotes the prior prediction, which is obtained from the $\chi_\mathrm{EFT}$ posterior for the pure nucleon matter and by choosing uniform priors on the eight additional parameters in our (extended) RMFT. The posterior-predictive band for the fit to $\chi_4^B/\chi_2^B$ alone is shown in blue whereas the orange band denotes the posterior band considering in addition the $\chi_{31}^{BS}/\chi_{11}^{BS}$ ratio. Both these bands match smoothly to the lattice QCD data at $\mu_B/T=2$, with the orange band lying very well within the $1\sigma$ band of the lattice data. The contour plots for the posterior distributions in the 16-dimensional space of the allowed baryon–meson couplings in our RMFT, constrained by these two susceptibility ratios, are shown in Figure~\ref{fig:corner_16d} in the Appendix.

\subsection{Neutron star matter and astrophysical consistency}
\label{sec:ns_structure}

Having constrained the hadronic model with available data from NEPs and lattice, to be as close to QCD as possible, we would now like to use this model to describe the neutron star matter. The hadronic matter inside a stable neutron star is cold, i.e., $T\ll \mu_B$ and is in $\beta$-equilibrium with electrons and muons. Thus, in order to describe the physics of neutron stars starting from our RMFT construction described in Eq.~\ref{eq:lagrangian} we need to additionally implement the $\beta$--equilibrium and charge neutrality conditions. This is implemented by ensuring local charge neutrality, i.e.,
\begin{equation}
\sum_b Q_b\, n_b - n_e - n_\mu = 0 .
\label{eq:charge_neutrality}
\end{equation}
where $n_e(n_\mu)$ are the number densities of electrons (muons), and $Q_b$ and $n_b$ are the charges and number density corresponding to a baryon labeled by $b$. Moreover, the chemical potential of each baryon species $b$ is written in terms of its baryon number $B_b$ as
\begin{equation}
\mu_b = B_b\,\mu_B - Q_b\,\mu_e ,
\label{eq:beta_equilibrium}
\end{equation}
where $\mu_e$ is the chemical potential for the electrons. Here the electrons and muons can be treated as free Fermi gases in equilibrium with each other i.e. $\mu_e = \mu_\mu$, each of which is populated once $\mu_e$ exceeds the corresponding rest mass of the electron or muon.

At a given baryon density $n_B = \sum_b n_b$, \, Eqs.~(\ref{eq:beta_equilibrium}) and (\ref{eq:charge_neutrality}) are solved simultaneously along with the equations of motion for the meson fields summarized in Eqs.~(\ref{eq:eom_sigma})--(\ref{eq:eom_phi}) for obtaining the mean field parameters along with $\mu_b$ and $\mu_e$. Below half the nuclear saturation density, $n_B < 0.08~\mathrm{fm}^{-3}$, the resulting $\beta$-equilibrated EOSs from our RMFT are matched with SLy4 EOS~\cite{DouchinHaensel_AA2001} for constructing the structural properties of neutron stars.

Recent observations of reasonably compact $2M_{\odot}$ neutron stars~\cite{Demorest_etal_nat2010, Antoniadis_sci2013} have led to intense investigation of the nature of QCD matter that might reside at their cores. It has been argued that such massive neutron star cores might contain a sizable amount of quark matter~\cite{Annala:2019puf} if the conformal bound for speed of sound, $c_s$, is not significantly violated. Our purely hadronic RMFT, even with the inclusion of the tower of excited baryon states and resonances, might not be able to account for this possibility. To address this, we therefore replace our microscopic RMFT with a model-agnostic, thermodynamically consistent sound-speed parametrization above a matching density, drawn from a uniform random distribution in the range $[3,\,4]\,n_{\rm sat}$, per sample. This is done to ensure that our final results do not depend on the density at which the matching is performed. We use ten different values of $c_s^2$ drawn uniformly in the range $[0,1]$ at densities up to $8\,n_{\rm sat}$, as control parameters and interpolate the values of baryon chemical potential, energy density and pressure as
\begin{equation}
\mu(n) = \mu(n_{\rm match})\,
\mathrm{e}^{\int_{n_{\rm match}}^{n}
\frac{c_s^2(n')}{n'}\,\mathrm{d}n'}~,
\label{eq:agnostic_cs2_mu}
\end{equation}
\begin{equation}
\varepsilon(n) = \varepsilon_{\rm match}
+ \int_{n_{\rm match}}^{n} \mu(n')\,\mathrm{d}n'~,p(n) = n\mu(n) - \varepsilon(n).
\label{eq:agnostic_cs2_epsilon}
\end{equation}
This method of interpolation is thermodynamically consistent and ensures causality is respected at all densities.

\section{Key findings and implications}
\label{sec:Implications}

\subsection{Improved constraints on baryon--meson couplings}

\begin{figure*}[!htbp]
\centering
\includegraphics[width=\textwidth]{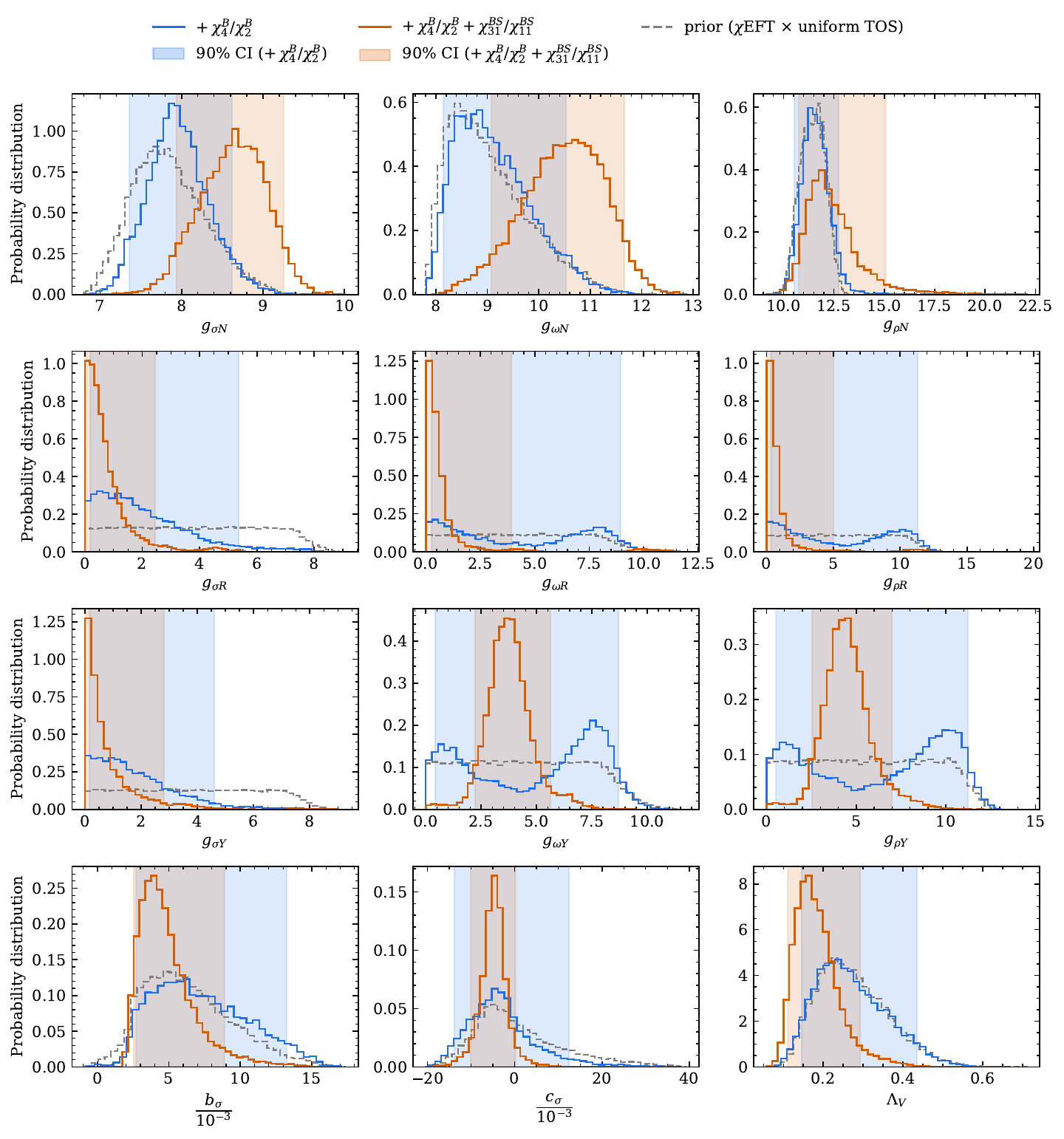}
\caption{
Constrained probability distributions for various baryon-meson couplings are illustrated in this figure. The \textit{top row} corresponds to the couplings between $\sigma$ (scalar--isoscalar), $\omega$ (vector--isoscalar), $\rho$ (vector--isovector) mesons and nucleons $N$ (protons or neutrons). The \textit{second row} corresponds to the couplings between each of the \{$\sigma$, $\omega$, $\rho$\} mesons and non-strange, non-nucleonic baryons $R$; and the \textit{third row} corresponds to the couplings between each of these mesons and strange baryons $Y$. Finally, the \textit{bottom row} corresponds to the probability distributions of the non-linear couplings for the meson self-interactions considered in this work. In each panel, the gray dashed histogram denotes the probability distribution of the corresponding coupling, arising from consistency with $\chi_\mathrm{EFT}$ only, which we use as the prior in the second stage of our Bayesian analysis. The solid histograms denote Bayesian posterior probability distributions after incorporating constraints from the baryon number susceptibility ratio $\frac{\chi_4^B}{\chi_2^B}(\hat{\mu}_B)$ alone (blue) and $\frac{\chi_4^B}{\chi_2^B}(\hat{\mu}_B)+\frac{\chi_{31}^{BS}}{\chi_{11}^{BS}}(\hat{\mu}_B)$ (orange), respectively. Furthermore, in each panel, the 90\% credible intervals corresponding to these two distributions obtained after imposing constraints from the respective susceptibility ratios are shown as shaded bands. 
}
\label{fig:BMcouplings}
\end{figure*}

In this section, we discuss the main highlights of our study. We show how constraining our RMFT parameters using inputs both from lattice QCD and previous inputs of NEPs from $\chi_\mathrm{EFT}$ provides improved constraints on the baryon-meson couplings. A statistical summary of the constraints on all baryon-meson couplings for the underlying RMFT is detailed in Table~\ref{tab:couplings_ci}. 

We first show our results for the inferred posterior probability distribution of the couplings of non-strange mesons to nucleons, which are summarized in panels in the top row of Figure~\ref{fig:BMcouplings}. Including just the lattice input for $\chi_4^B/\chi_2^B$ does not allow us to further constrain the values of coupling for the interaction due to $\sigma$ and $\omega$-mesons as compared to using just the NEPs as priors. However, additionally including the baryon--strangeness correlation ratio from lattice QCD  shifts the most probable values of these couplings to noticeably larger magnitudes than previously estimated. The shift in the probability distribution of the repulsive $\rho$-nucleon coupling due to this additional input from lattice QCD is barely visible.

This additional input also impacts the predictions for the couplings of mesons to non-strange baryons other than the nucleons. This is illustrated in panels in the second row of Figure~\ref{fig:BMcouplings}. With the inclusion of the lattice data on the baryon--strangeness correlation ratio along with $\chi_4^B/\chi_2^B$, it is possible to estimate upper bounds on the couplings with high accuracy.  

We next show the efficacy of our procedure for constraining the allowed range of couplings of non-strange mesons to strange baryons $Y$, results for which are summarized in the third-row panels of Figure~\ref{fig:BMcouplings}. When we consider only the ratio of baryon number susceptibilities, the posterior distributions for $g_{\omega Y}$ and $g_{\rho Y}$ are broad and exhibit weak bimodality. These constraints become distinctly prominent upon additionally including lattice QCD data for the ratio of baryon--strangeness correlations. The inferred $90\%$ credible interval for $g_{\omega Y}$ is $3.72 \ [2.22, 5.63]$\,, and for $g_{\rho Y}$ is $4.41 \ [2.53, 7.00]$. Our multivariate analysis thus allows us to place more stringent bounds on the repulsive couplings involving the $\rho$ and $\omega$ mesons. On the other hand, the coupling between the $\sigma$ meson and strange baryons remains heavily skewed toward zero, yielding only an upper bound of $2.82$ at a $90\%$ credible level. The combined constraint thus points toward an effectively strong repulsive potential for strange baryons in dense hadronic matter. Constraining the attractive interactions in this sector will require additional input of susceptibility ratios from lattice QCD.

\begin{figure*}[htbp]
\centering
\includegraphics[width=0.76\textwidth]{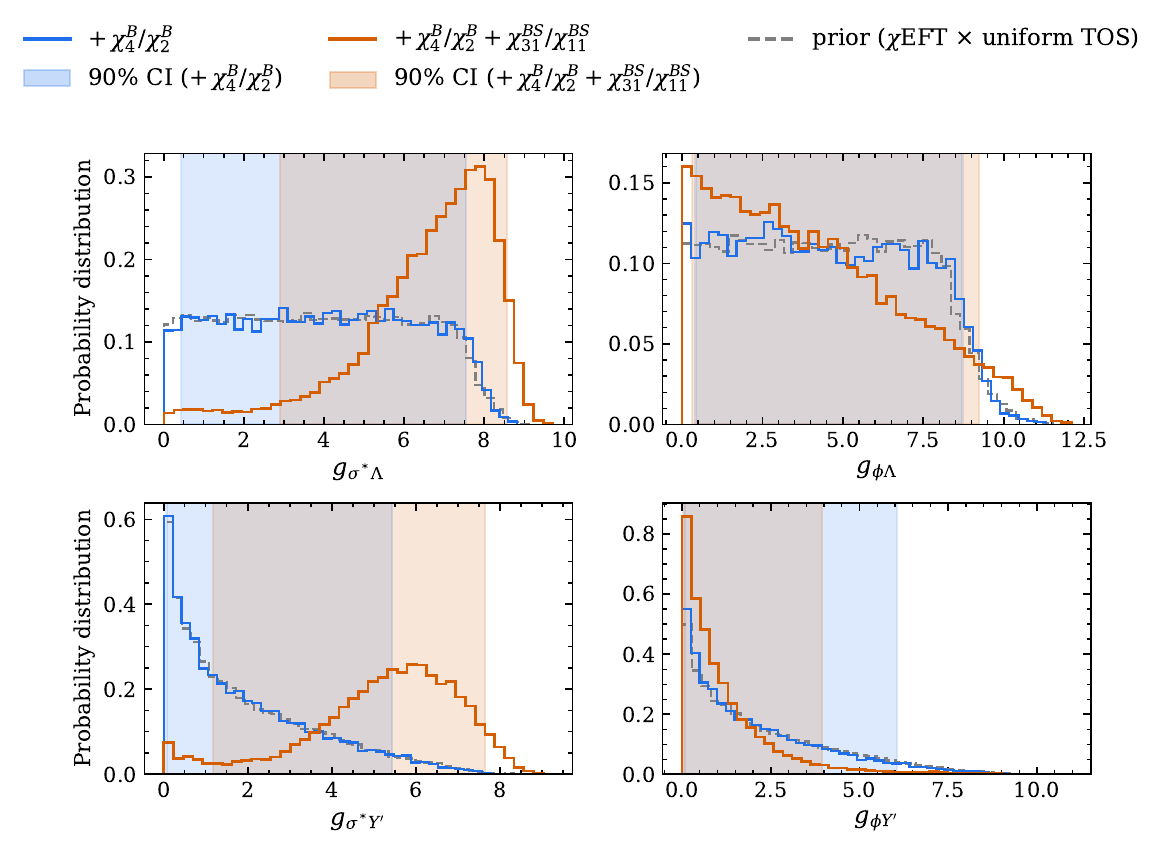}
\caption{This figure is similar to Figure~\ref{fig:BMcouplings}, but demonstrates the constraints on hidden strange sector couplings. The corresponding prior and Bayesian posterior probability distributions (same notations as Figure~\ref{fig:BMcouplings}) for couplings of the scalar and vector hidden-strange mesons, $\sigma^*$ and $\phi$, respectively, with the $\Lambda$ hyperon are shown in the \emph{top panel}. The corresponding couplings to other strange baryons are shown in the \emph{bottom panel}.
}
\label{fig:BSMcouplings}
\end{figure*}

We also study the coupling of the strange mesons to $\Lambda$-hyperon alone and to all other heavier strange baryons and resonances; the probability distribution of the corresponding couplings are shown in Figure~\ref{fig:BSMcouplings}. With our present strategy we cannot constrain the coupling of the $\Lambda$ hyperon to strange vector mesons, whereas including the baryon--strangeness correlation data from lattice QCD allows us to get an improved constraint on the scalar couplings. For the excited baryon states too, we can place tighter bounds for the scalar coupling with the baryon--strangeness correlation data, whereas we can provide only an upper bound on the coupling to the vector meson $\phi$.  This is quite unlike the trend that we observed for the couplings of the strange baryons to non-strange mesons, where scalar interactions were weaker compared to the vector interactions.

The results for the probability distributions of the three different non-linear couplings are summarized in the three panels in the bottom row of Figure~\ref{fig:BMcouplings}. Incorporating additional input from the baryon number–strangeness correlation ratio enables more stringent constraints on these non-linear couplings than what we could have obtained from the lattice QCD ratio $\chi_4^B/\chi_2^B$ alone. This is particularly crucial for the $\Lambda_V$ coupling---which affects the slope parameter of the symmetry energy and the neutron-skin thickness---for which the probability distribution peaks at a lower value than previously constrained from the NEPs.

\begin{figure*}[!htbp]
    \centering
    \includegraphics[width=\textwidth, height=0.65\textwidth]{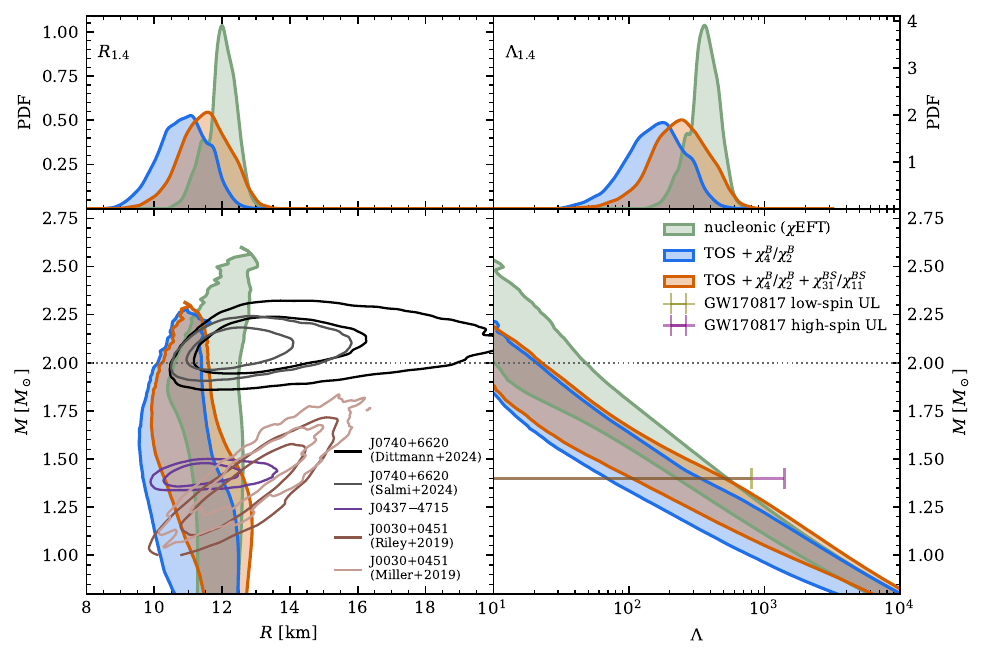}
    \caption{
    This figure illustrates the radius $R$ (left panels) and tidal deformability $\Lambda$ (right panels) as a function of the gravitational mass $M$ of neutron stars for our RMFT, with the nucleonic sector supplemented with the tower of states of baryons and resonances. The shaded regions represent $90\%$ credible bands evaluated at a fixed $M$ on the stable branch. In each panel, the green shaded region is the data from purely nucleonic RMFT constrained by $\chi_\mathrm{EFT}$ alone (Fig.~\ref{fig:EbyA_pnm}), without the tower of states; the blue and orange shaded regions include the tower of states, constrained additionally by $\chi_4^B/\chi_2^B(\hat{\mu}_B)$ and by $\frac{\chi_4^B}{\chi_2^B}(\hat{\mu}_B)+\frac{\chi_{31}^{BS}}{\chi_{11}^{BS}}(\hat{\mu}_B)$, respectively. For the blue and orange bands, above a matching density drawn uniformly from $[3,4]\,n_{\rm sat}$, the RMFT is replaced by a model-agnostic, thermodynamically consistent and causal sound-speed parametrization to account for physics beyond RMFT. The upper panels show the corresponding posterior distributions of $R_{1.4}$ and $\Lambda_{1.4}$ for a $1.4\,M_\odot$ neutron star, having the same abscissa as the panels beneath them. Mass--radius constraints for PSR J0740+6620 from NICER and XMM~\cite{Dittmann_2024, dittmann_2024_10215109_Zenodo, Salmi_2024, salmi_2024_10519473_Zenodo}, PSR J0437–4715~\cite{Choudhury_2024, choudhury_2024_12703175_Zenodo}, and PSR J0030–0451 from NICER~\cite{Miller_2019, miller_2019_3473466_Zenodo, Riley_2019, riley_2022_7096789_Zenodo} are overlaid, together with the GW170817 upper limits on the dimensionless tidal deformability parameter for a $1.4\,M_\odot$ neutron star under the inferred low-spin ($\Lambda_{1.4} < 800$, dashed) and high-spin ($\Lambda_{1.4} < 1400$, dotted) priors~\cite{GW170817_PRL2017}. The dotted horizontal line denotes the observed $2\,M_\odot$ neutron star limit~\cite{Demorest_etal_nat2010, Antoniadis_sci2013}.
    }
    \label{fig:MRLam_lQCD_agnostichd}
\end{figure*}

\subsection{Implications for the physics of neutron stars}

Having constrained the parameter space of our RMFT couplings, we now discuss their implications for the dense matter EOS of cold neutron star cores in beta equilibrium. The purely nucleonic RMFT, constrained by $\chi_\mathrm{EFT}$ alone, provides a radius $R_{1.4}=12.00^{+0.61}_{-0.86}~\mathrm{km}$ and a tidal deformability $\Lambda_{1.4}=368^{+163}_{-143}$ for a $1.4\,M_\odot$ neutron star within $90\%$ credible intervals. The probability distributions of the radius and tidal deformability for a $1.4M_{\odot}$ neutron star are shown in the upper panels of Figure~\ref{fig:MRLam_lQCD_agnostichd}. In the lower-right panel, we show the allowed parameter space obtained in our analysis in the mass-tidal deformability ($M-\Lambda$) plane. The predicted values of $\Lambda$ are consistent with the GW170817 upper bounds for both the low- and high-spin priors~\cite{GW170817_PRL2017}. Furthermore, the allowed region of  radii, as shown in the lower-left panel of Figure~\ref{fig:MRLam_lQCD_agnostichd}, is compatible with the NICER observations of the $\sim 1.4\,M_\odot$ pulsars PSR J0437$-$4715~\cite{Choudhury_2024, choudhury_2024_12703175_Zenodo} and PSR J0030–0451~\cite{Miller_2019, miller_2019_3473466_Zenodo, Riley_2019, riley_2022_7096789_Zenodo} as well as with PSR J0740$+$6620~\cite{Dittmann_2024, dittmann_2024_10215109_Zenodo, Salmi_2024, salmi_2024_10519473_Zenodo, Miller_2021, Riley_2021} probing a massive  $2M_{\odot}$ neutron star. For each NICER source, we show the posterior of the headline model of the corresponding reference.

More interesting mass-radius constraints arise when we include the tower of states of excited baryons and resonances. This is summarized in the lower-left panel of Figure~\ref{fig:MRLam_lQCD_agnostichd}. We recall here that in order to accommodate unconsidered physical effects, e.g., phase transitions, the RMFT characterization of the speed of sound is replaced above a matching density of $3$--$4\,n_{\rm sat}$ by a model-agnostic, thermodynamically consistent and causal parametrization, which is discussed in detail in Section~\ref{sec:ns_structure}. The mass-radius values within our RMFT, with couplings obtained from a joint fit to the lattice QCD data on $\chi_4^B/\chi_2^B$ and $\chi_{31}^{BS}/\chi_{11}^{BS}$, are shown as an orange region in the lower-left panel of Figure~\ref{fig:MRLam_lQCD_agnostichd}. 

Constraining the couplings with $\chi_4^B/\chi_2^B$ alone (blue) yields more compact stars, with $R_{1.4}=10.89^{+1.08}_{-1.22}$~km and $\Lambda_{1.4}=164^{+173}_{-95}$. However, while using couplings constrained by the joint fit to $\chi_4^B/\chi_2^B$ and $\chi_{31}^{BS}/\chi_{11}^{BS}$ (orange), we obtain $R_{1.4}=11.52^{+1.11}_{-1.22}$~km and $\Lambda_{1.4}=245^{+260}_{-140}$ for a $1.4\,M_\odot$ star. While both ensembles are consistent with the NICER mass--radius measurements and with the GW170817 bound on the tidal deformability parameter, the joint fit---which shifts the band towards larger radii---overlaps more closely with the NICER observations.

\section{Summary \& Outlook}

In this work, we have made notable progress toward achieving a more realistic hadrodynamic description of the confined phase of QCD.  Our starting point was a relativistic mean-field description of interacting hadronic matter that includes, besides nucleons, a tower of baryons and resonances up to $3~\mathrm{GeV}$, which is well motivated by QCD. We have now constrained various couplings using controlled \textit{ab initio} inputs from two different and  complementary parameter regimes. First, the nucleon sector was constrained from the $\chi_\mathrm{EFT}$ energy per nucleon of pure neutron matter, using a likelihood that accounts for the strong correlation of the theoretical uncertainty across a range of baryon densities. We next used the resulting posteriors as priors for constraining the couplings of the excited baryons by incorporating two lattice-QCD susceptibility ratios at $T=135~\mathrm{MeV}$ and finite $\hat{\mu}_B \in [2.0, 3.0]$.

Though we could achieve stronger constraints on the isovector properties of cold, neutron-rich matter near saturation density, namely the symmetry energy and its slope, together with the nucleon effective mass using $\chi_\mathrm{EFT}$ input, the saturation properties of symmetric nuclear matter remain essentially unconstrained. On the other hand, input from lattice QCD in comparatively hot nuclear matter at lower baryon densities allowed us to constrain the interactions due to the higher excited states in the baryon sector that have negligible contributions to the NEPs in cold nuclear matter. Whereas the ratio $\chi_4^B/\chi_2^B$ from lattice QCD alone constrains only repulsive interactions,  adding the baryon--strangeness correlation ratio $\chi_{31}^{BS}/\chi_{11}^{BS}$ provides more detailed information about the hadron interactions. The non-strange baryon resonances are found to be weakly interacting. For strange baryons, the couplings to the non-strange vector mesons come out to be somewhat larger than their couplings to scalar mesons, whereas for the hidden strange mesons the opposite trend is observed. However, note that our present construction assumes that the couplings of all the strange baryons in the tower of states with the ground state mesons are the same. 

We also investigated the implications of our findings for understanding nuclear matter in neutron stars across different mass–radius limits. The probability distributions of the resulting radii and tidal deformabilities remain consistent with different astrophysical observations, both when calculated within the nucleonic RMFT alone, constrained from pure neutron-matter properties, and when the RMFT is augmented with excited states of baryons and resonances whose couplings are constrained with the lattice QCD data.  However, including the contribution of the tower of excited baryon states tends to favor more compact neutron stars. In order to understand the implications of these improved bounds on the baryon-meson couplings, one has to include them in the full dynamical modeling of neutron stars and their mergers. Moreover, one would expect that these excited baryon states will leave discernible imprints on the transport properties such as thermal conductivity, bulk viscosity, etc., that are relevant for neutron-star mergers~\cite{Alford_PRL2018, Alford_PRC2019, Alford_universe2021},  and therefore would provide a natural test of the constrained couplings. We note here that the ground-state mesons such as pions and kaons are not included in our RMFT construction; however, some of the excited states arising out of pion--nucleon attractive interactions are included in the tower of states. A natural extension of our work will involve including these pseudo-Goldstone modes due to chiral symmetry breaking and constraining their interaction couplings to study their implications for neutron star mergers.

\acknowledgements 

The authors gratefully acknowledge Aman Abhishek for discussions and his assistance during the early stages of this work. A.M. acknowledges support from the DST-SERB Start-up Research Grant No. SRG/2020/001290. S.S. acknowledges support from the Anusandhan National Research Foundation, Government of India, under Project ID ANRF/ARG/2025/010416/PS.

\appendix

\section*{Appendix}

\section{Reconstructing input from lattice QCD}

We tabulate the lattice data for different higher-order baryon number and baryon--strangeness susceptibilities at $\hat{\mu}_B=0$ and $T=135~\mathrm{MeV}$, obtained from Ref.~\cite{Borsanyi_2018}, in Table~\ref{tab:lQCD_sus_values}. Using these data, we reconstruct two different susceptibility ratios at finite $\hat{\mu}_B$ values that enter into the likelihood analysis summarized in Eqs.~(\ref{eq:chi4B_by_chi2B})--(\ref{eq:chi31BS_by_chi11BS}), using a Taylor-series expansion. In the following two sections, we discuss various systematics involved in the truncation of this series expansion and the choice for $\hat{\mu}_B$ values where we perform this reconstruction.

\begin{table*}[!h]
\centering
\caption{Susceptibilities at zero baryon chemical potential and $T = 135~\mathrm{MeV}$ taken from Ref.~\cite{Borsanyi_2018}. The last row gives the resolution parameter at the $k$-th order, $z_k = |c_k|/e_k$. Coefficients with $z_k < 1$ (i.e., $\chi_8^B$ and $\chi_{71}^{BS}$) are noise-dominated and dropped from the Taylor series.}
\label{tab:lQCD_sus_values}
\resizebox{0.75\textwidth}{!}{%
\begin{tabular}{|c|c|c|c|c|c|c|c|c|c|c|c|c|}
\toprule
Parameter & $\chi_2^B$ & $\chi_4^B$ & $\chi_6^B$ & $\chi_8^B$ & $\chi_{11}^{BS}$ & $\chi_{31}^{BS}$ & $\chi_{51}^{BS}$ & $\chi_{71}^{BS}$ & $\chi_2^S$ & $\chi_4^S$ & $\chi_{22}^{BS}$ & $\chi_{42}^{BS}$ \\
\midrule
Mean value & $0.034$ & $0.035$ & $0.035$ & $0.012$ & $-0.014$ & $-0.015$ & $-0.013$ & $0.000$ & $0.110$ & $0.146$ & $0.022$ & $0.023$ \\ \hline
Error & $0.000$ & $0.003$ & $0.014$ & $0.050$ & $0.000$ & $0.001$ & $0.006$ & $0.018$ & $0.000$ & $0.000$ & $0.002$ & $0.006$ \\
\midrule
$z_k=|c_k|/e_k$ & $-$ & $11.7$ & $2.5$ & $0.2$ & $-$ & $15.0$ & $2.2$ & $0.0$ & $-$ & $-$ & $11.0$ & $3.8$ \\
\bottomrule
\end{tabular}
}
\end{table*}

\subsection{Truncation of the Taylor series }
\label{app:truncation}

We discuss here the systematic errors involved in the truncation of the Taylor series of different baryon-number and baryon--strangeness susceptibilities. The order $k$ at which each series is truncated depends on the error on the coefficient $c_k$ appearing in front of the term $\hat{\mu}_B^k$. If the uncertainty on this coefficient is $e_k$, then this term is retained only if it is resolved above its own uncertainty, i.e., the resolution parameter $z_k$ satisfies 
\begin{equation}
\label{eq:zres}
z_k = \frac{|c_k|}{e_k} \gtrsim 1 .
\end{equation}

The Taylor series of the quantities $\chi_2^B(\hat{\mu}_B)$ and $\chi_4^B(\hat{\mu}_B)$ involve coefficients up to $\chi_6^B(\hat{\mu}_B=0)$, which has $z=2.5$. For the baryon--strangeness correlations, the Taylor series involves coefficients up to $\chi_{51}^{BS}(\hat{\mu}_B=0)$, which has a resolution $z=2.2$. The next higher-order Taylor coefficients, $\chi_8^B$ (for baryon-number fluctuations) and $\chi_{71}^{BS}$ (for baryon--strangeness correlations) have resolutions of $z\simeq 0.2 \,\rm{ and }\, 0$, respectively, and are therefore not considered. Including coefficients that are consistent with zero and dominated by large uncertainties would propagate large errors into the Taylor series without providing additional information.

\begin{figure*}[!htbp]
\centering
\includegraphics[width=0.9\textwidth]{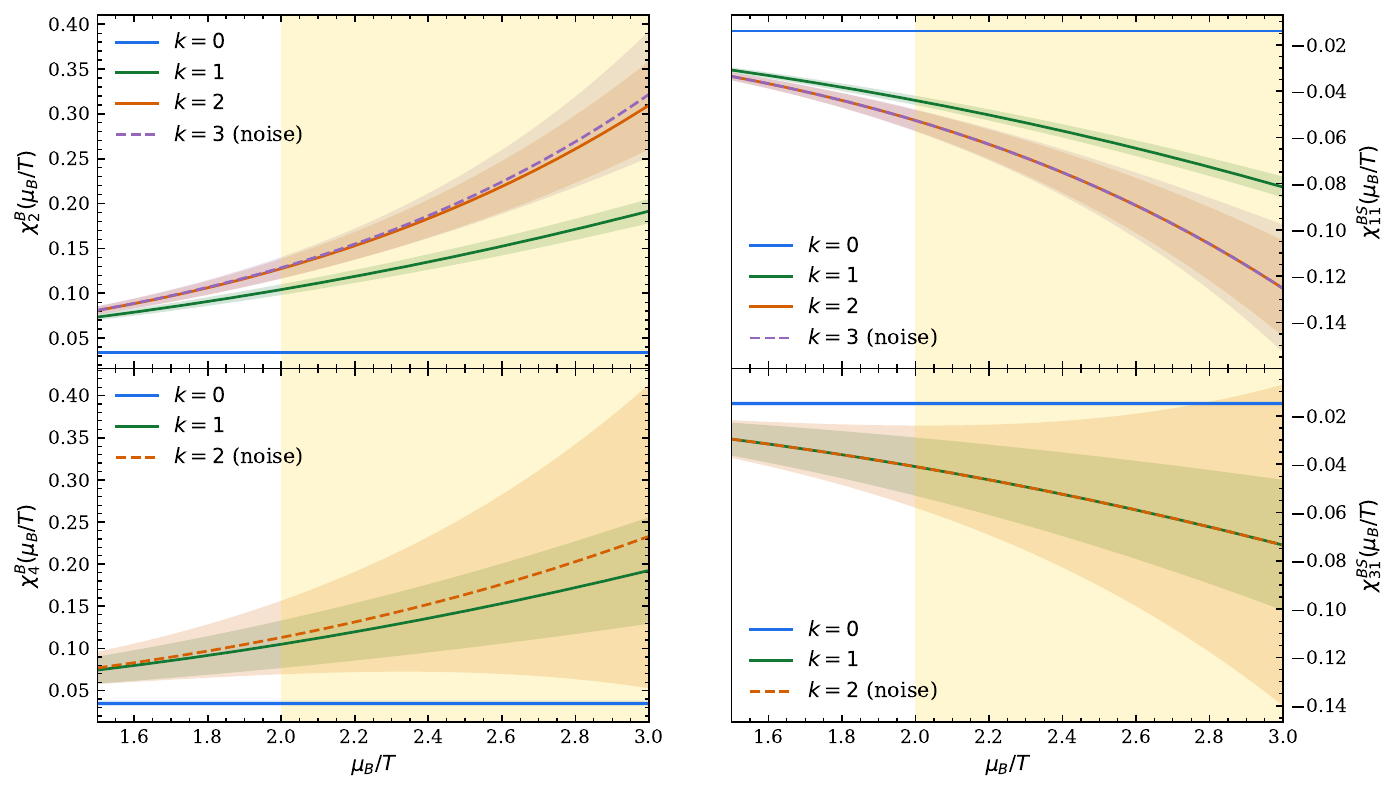}

\caption{
This figure presents the Taylor series reconstruction of $\chi_2^B(\hat{\mu}_B)$ and $\chi_4^B(\hat{\mu}_B)$ (\textit{left panel}) and of $\chi_{11}^{BS}(\hat{\mu}_B)$ and $\chi_{31}^{BS}(\hat{\mu}_B)$ (\textit{right panel}), at $T=135~\mathrm{MeV}$ as a function of $\hat{\mu}_B$ and truncated at $\mathcal{O}(\hat{\mu}_B^{2k})$. Solid bands denote $1\sigma$ uncertainties of the series truncated at each order, with the central values of the included orders indicated by solid lines and the discarded noise-dominated order by a dashed line. The yellow region indicates the window $\hat{\mu}_B \in [2.0,3.0]$ in which this reconstruction was performed. 
}
\label{fig:recon}
\end{figure*}

Figure~\ref{fig:recon} shows the error bands associated with the Taylor series truncated at $\mathcal{O}(\hat{\mu}_B^{2k})$ of different susceptibilities for $\chi_2^B(\hat{\mu}_B)$, $\chi_4^B(\hat{\mu}_B)$, $\chi_{11}^{BS}(\hat{\mu}_B)$, and $\chi_{31}^{BS}(\hat{\mu}_B)$. The yellow region denotes the range of $\hat{\mu}_B \in [2.0,3.0]$ over which the lattice susceptibility ratios $\chi_4^B/\chi_2^B(\hat{\mu}_B)$ and $\chi_{31}^{BS}/\chi_{11}^{BS}(\hat{\mu}_B)$ are reconstructed for our likelihood analysis. We observe that for all these observables, the error band due to the noisiest Taylor coefficient widens sharply at larger values of $\hat{\mu}_B$. This justifies our methodology to truncate at a lower order as retaining these terms would degrade rather than improve the constraint on the reconstructed lattice data. In the future, with improvement in lattice data, we expect that these higher-order coefficients will be available with controllable error, which will allow us to extend the window to larger $\hat{\mu}_B$ for our likelihood analysis.

\begin{figure*}[htbp]
    \includegraphics[width=0.84\columnwidth]{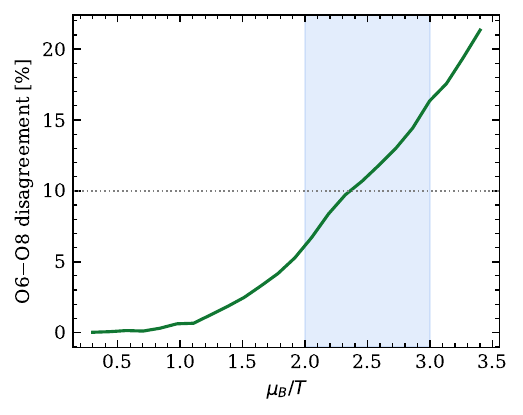}\label{subfig:lattice_validity}
    \includegraphics[width=0.84\columnwidth]{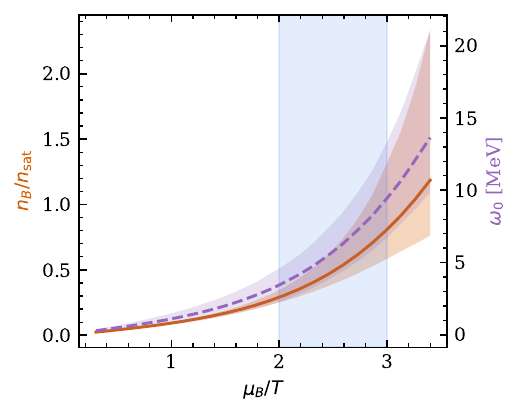}\label{subfig:mean_fields_prior} \\
    \caption{The \textit{left panel} presents the relative percentage of truncation errors at $\mathcal{O}(\hat{\mu}_B^6)$ versus $\mathcal{O}(\hat{\mu}_B^8)$ in the Taylor series of $\chi_4^B/\chi_2^B(\hat{\mu}_B)$ as a function of $\hat{\mu}_B$. The horizontal dotted line marks the adopted $10\%$ truncation tolerance. In the \textit{right panel}, the corresponding $90\%$ credible bands for priors on the reduced baryon density $n_B/n_{\rm sat}$ and mean-field of the vector meson $\omega_0$ are shown as red and magenta bands, respectively. In both panels, the shaded vertical band denotes the adopted fitting window, $\hat{\mu}_B\in[2.0,\,3.0]$; for details, see text in Sec.~\ref{app:window}.}
\label{fig:window_diagnostics}
\end{figure*}

\subsection{Setting the range of $\hat{\mu}_B$ for the input lattice data}
\label{app:window}

In this section, we show how we obtain susceptibility ratios, using a Taylor-series expansion, in the range of $\hat{\mu}_B \in [2.0, 3.0]$ at $T=135~\mathrm{MeV}$. We now motivate our choice of this particular $\hat{\mu}_B$ window. On the one hand, the lower boundary of this window is chosen such that a finite non-trivial mean-field value of $\omega_0$ is obtained; on the other hand, the upper boundary is set by the availability of reliable calculations from lattice QCD. This allows the baryon-number susceptibilities obtained from the RMFT to be sensitive to the underlying baryon--meson couplings.

Truncated Taylor-series expansions of the cumulant ratios are reliable only within specific ranges of $\hat{\mu}_B$. We have already quantified the truncation errors in the last section. Here we show the ratio of truncation errors in the left panel of Figure~\ref{fig:window_diagnostics} while extending the expansion from $\mathcal{O}(\hat{\mu}_B^6)$ to $\mathcal{O}(\hat{\mu}_B^8)$ for $\chi_4^B/\chi_2^B(\hat{\mu}_B)$. The ratio is $\sim 10\%$ near $\hat{\mu}_B\simeq 2.4$ and reaches $\simeq~15\%$ at $\hat{\mu}_B=3.0$. Beyond this, since the truncation error grows further, we choose $\hat{\mu}_B=3.0$ as the largest value where we reconstruct the lattice data using a Taylor-series expansion.

To justify our choice for the lower end of the window in $\hat{\mu}_B$, we note that at $\hat{\mu}_B=0$ the vector (repulsive) mean field vanishes and the cumulant ratios carry no information about baryon--meson couplings. Repulsive interactions become dominant only as the baryon density grows. By sampling the couplings of the ground-state mesons with the tower-of-states baryons from their priors and the nucleonic sector from the $\chi_\mathrm{EFT}$ posterior, we obtain the resulting $90\%$ prior band for the reduced baryon density $n_B/n_{\rm sat}$ (marked as the red band in the right panel of Figure~\ref{fig:window_diagnostics}). It is evident that for $\hat{\mu}_B < 3$, the system remains in the sub-saturation density region. In this region, the mean-field value of the $\omega_0$ field grows, which is shown by the dotted magenta line (median) and band ($90\%$ prior) in the same Figure. For $\hat{\mu}_B < 2$, the mean-field value of $\omega_0$, and thus the repulsive interactions among the baryons remain negligibly small. Therefore, the susceptibility ratios are largely insensitive to the couplings below $\hat{\mu}_B=2$, which sets the choice of the lower bound of the window in our analysis.

The two edges of our window in $\hat{\mu}_B$ (see Figure~\ref{fig:window_diagnostics}) balance two competing factors. While the Taylor-series reconstruction of the lattice data is more reliable at the lower edge of our window, the sensitivity to baryon interactions becomes noticeable at higher values of $\hat{\mu}_B$. 

\begin{figure*}[hbt!]
    \centering
    \includegraphics[clip,width=0.9\textwidth]{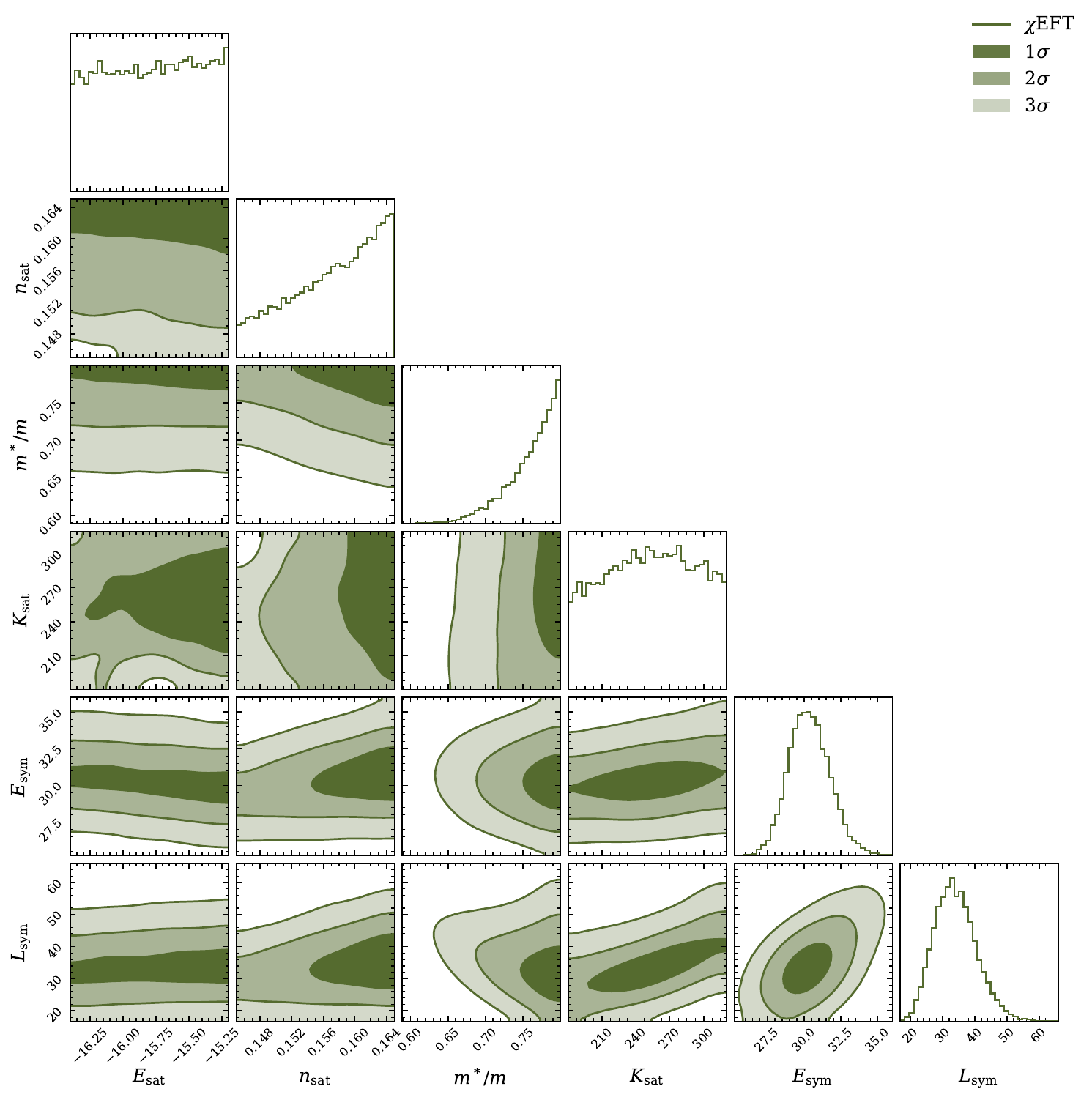}
    \caption{This figure illustrates the full Bayesian posterior distributions of the six-dimensional nuclear empirical parameters $\{E_{\rm sat},\, n_{\rm sat},\, m^*/m,\, K_{\rm sat},\, E_{\rm sym},\, L_{\rm sym}\}$, constrained by the $\chi_\mathrm{EFT}$ input of the energy per neutron for pure neutron matter. The detailed methodology is discussed in subsection~\ref{subsec:constraints_chiEFT}.}
    \label{fig:corner_chieft_6d}
\end{figure*}

\section{Bayesian Posterior Distributions}
\label{app:posterior_distr}

\begin{figure*}[hbt!]
    \centering
    \includegraphics[clip,width=\textwidth]{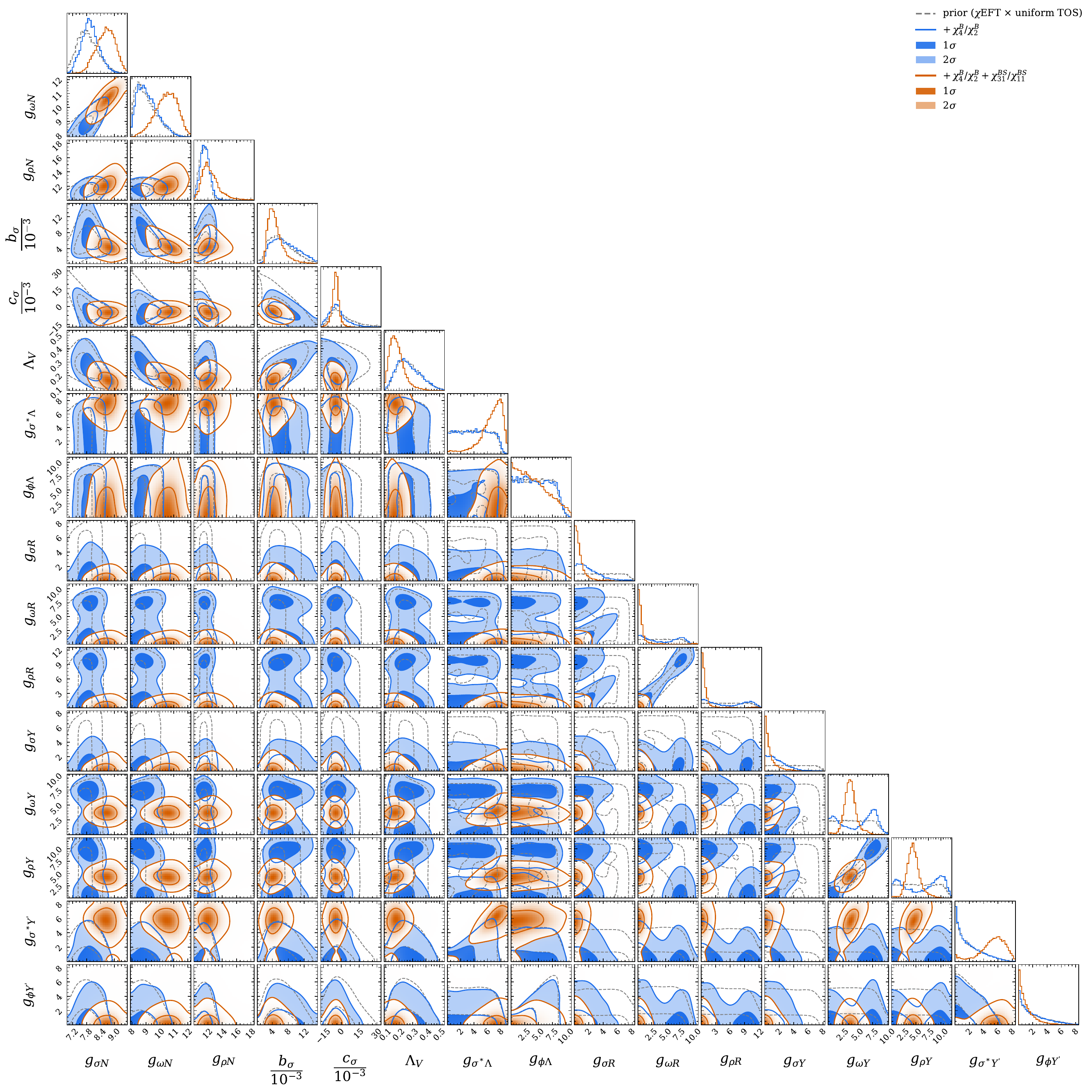}
    \caption{The full Bayesian posterior distributions for the 16-dimensional parameter space of baryon--meson couplings are presented in this corner plot. The first six parameters are the nucleonic couplings $g_{\sigma N},\, g_{\omega N},\, g_{\rho N},\, b_\sigma,\, c_\sigma,\, \Lambda_V$. The remaining ten are the couplings with the tower of states of baryons. The couplings $g_{\sigma^*\Lambda},\, g_{\phi\Lambda}$ refer to the interaction of the $\Lambda$ hyperon with hidden-strange mesons. The ground-state non-strange mesons couple to non-strange, non-nucleonic baryons through $g_{\sigma R}$, $g_{\omega R}$, and $g_{\rho R}$, and to strange baryons through $g_{\sigma Y}$, $g_{\omega Y}$, and $g_{\rho Y}$, whereas hidden-strange mesons couple to strange baryons other than the $\Lambda$ through $g_{\sigma^* Y^{'}},\, g_{\phi Y^{'}}$. The priors, shown as gray dashed contours, combine the $\chi_\mathrm{EFT}$ posterior for the nucleonic sector with uniform priors on the eight tower-of-states parameters ($g_{\sigma^*\Lambda}/g_{\sigma N}$, $g_{\phi\Lambda}/g_{\omega N},\alpha^{(\rm at)},\alpha^{(\rm r)},\alpha^{(\rm at)}_{\rm ns},\alpha^{(\rm r)}_{\rm ns},\alpha^{(\rm at)}_{\rm s},\alpha^{(\rm r)}_{\rm s}$), which are then used for sampling in Bayesian parameter estimation. These are compared with the posteriors constrained using lattice QCD data with $\frac{\chi_4^B}{\chi_2^B}(\hat{\mu}_B)$ alone (shown as blue contours), and with $\frac{\chi_4^B}{\chi_2^B}(\hat{\mu}_B)+\frac{\chi_{31}^{BS}}{\chi_{11}^{BS}}(\hat{\mu}_B)$ (shown as orange contours). The detailed methodology is discussed in subsection~\ref{subsec:constraints_lQCD}.
    }\label{fig:corner_16d}
\end{figure*}

In this section, we summarize the results for different posterior distributions obtained in our likelihood analysis. The posterior distributions of the nuclear empirical parameters constrained by $\chi_\mathrm{EFT}$ are compiled in Figure~\ref{fig:corner_chieft_6d}. In Figure~\ref{fig:corner_16d}, we show the posterior distributions of different baryon--meson couplings with combined inputs from both $\chi_\mathrm{EFT}$ and lattice QCD. Table~\ref{tab:couplings_ci} summarizes the corresponding one-dimensional marginal distributions obtained in our analysis. The detailed methodology is discussed in Section~\ref{sec:constraints}.

\begin{table*}[htbt]
\centering
\caption{Posterior medians and equal-tailed credible intervals for different baryon--meson couplings obtained from our likelihood analysis. For parameters whose probability distributions are peaked at zero, only upper limits are quoted. For these couplings, the median values and the lower bounds of the credible intervals are denoted by crosses (`$\times$'). The detailed methodology is discussed in Section~\ref{sec:constraints}.}
\label{tab:couplings_ci}
\begin{tabular}{l|c|c|c|c|c|c}
\toprule
Parameter & Median & 50\% & 90\% & $1\sigma$ & $2\sigma$ & $3\sigma$ \\
\midrule
\multicolumn{7}{l}{$\chi$EFT (nucleonic RMFT)} \\
\midrule
$g_{\sigma N}$ & 7.81 & [7.52, 8.14] & [7.18, 8.66] & [7.40, 8.31] & [7.07, 8.84] & [6.89, 9.26] \\
$g_{\omega N}$ & 8.89 & [8.44, 9.51] & [8.05, 10.5] & [8.28, 9.86] & [7.96, 10.9] & [7.85, 11.7] \\
$g_{\rho N}$ & 11.5 & [11.0, 11.9] & [10.5, 12.4] & [10.8, 12.1] & [10.3, 12.6] & [9.96, 13.0] \\
$b_\sigma/10^{-3}$ & 5.94 & [3.97, 8.31] & [2.10, 12.1] & [3.23, 9.65] & [1.30, 13.5] & [-0.272, 15.9] \\
$c_\sigma/10^{-3}$ & -1.25 & [-6.20, 6.46] & [-12.2, 21.9] & [-8.16, 11.3] & [-14.3, 27.2] & [-18.0, 36.4] \\
$\Lambda_V$ & 0.264 & [0.209, 0.332] & [0.149, 0.437] & [0.187, 0.365] & [0.134, 0.476] & [0.101, 0.580] \\
\midrule
\multicolumn{7}{l}{$\chi$EFT $+\,\chi_4^B/\chi_2^B$} \\
\midrule
$g_{\sigma N}$ & 7.94 & [7.71, 8.19] & [7.36, 8.62] & [7.59, 8.33] & [7.27, 8.76] & [6.97, 9.15] \\
$g_{\omega N}$ & 9.02 & [8.56, 9.59] & [8.14, 10.5] & [8.40, 9.90] & [8.05, 10.8] & [7.87, 11.7] \\
$g_{\rho N}$ & 11.6 & [11.1, 12.0] & [10.5, 12.7] & [10.9, 12.2] & [10.3, 13.1] & [9.76, 15.8] \\
$b_\sigma/10^{-3}$ & 6.81 & [4.68, 9.55] & [2.71, 13.2] & [3.83, 10.9] & [2.28, 14.2] & [-0.0154, 16.1] \\
$c_\sigma/10^{-3}$ & -3.95 & [-8.19, 0.558] & [-13.9, 12.5] & [-10.3, 4.02] & [-15.4, 17.8] & [-18.5, 32.8] \\
$\Lambda_V$ & 0.259 & [0.204, 0.331] & [0.145, 0.436] & [0.183, 0.367] & [0.128, 0.470] & [0.0940, 0.560] \\
$g_{\sigma^* \Lambda}$ & 3.98 & [2.01, 5.88] & [0.436, 7.54] & [1.28, 6.63] & [0.205, 7.83] & [0.0127, 8.57] \\
$g_{\phi \Lambda}$ & 4.37 & [2.21, 6.73] & [0.419, 8.72] & [1.38, 7.59] & [0.181, 9.18] & [0.00887, 10.6] \\
$g_{\sigma R}$ & 1.68 & [0.822, 2.90] & [0.185, 5.37] & [0.531, 3.58] & [0.0908, 6.56] & [0.00678, 7.90] \\
$g_{\omega R}$ & 4.31 & [1.35, 7.48] & [0.257, 8.95] & [0.797, 8.06] & [0.120, 9.41] & [0.00960, 10.8] \\
$g_{\rho R}$ & 5.42 & [1.74, 9.47] & [0.332, 11.3] & [1.02, 10.3] & [0.158, 11.8] & [0.0112, 14.3] \\
$g_{\sigma Y}$ & 1.49 & [0.728, 2.56] & [0.145, 4.61] & [0.456, 3.20] & [0.0667, 5.74] & [0.00446, 7.79] \\
$g_{\omega Y}$ & 5.84 & [1.90, 7.53] & [0.411, 8.71] & [1.15, 7.98] & [0.213, 9.10] & [0.0131, 10.2] \\
$g_{\rho Y}$ & 7.31 & [2.44, 9.76] & [0.525, 11.2] & [1.44, 10.4] & [0.266, 11.6] & [0.0157, 12.5] \\
$g_{\sigma^* Y^{'}}$ & 1.45 & [0.512, 2.97] & [0.0639, 5.44] & [0.284, 3.82] & [0.0261, 6.18] & [0.00142, 7.44] \\
$g_{\phi Y^{'}}$ & $\times$ & [$\times$, 3.16] & [$\times$, 6.05] & [$\times$, 4.11] & [$\times$, 7.00] & [$\times$, 8.96] \\
\midrule
\multicolumn{7}{l}{$\chi$EFT $+\,\chi_4^B/\chi_2^B+\chi_{31}^{BS}/\chi_{11}^{BS}$} \\
\midrule
$g_{\sigma N}$ & 8.65 & [8.35, 8.92] & [7.94, 9.26] & [8.21, 9.04] & [7.80, 9.37] & [7.32, 9.74] \\
$g_{\omega N}$ & 10.5 & [9.94, 11.1] & [9.08, 11.7] & [9.66, 11.3] & [8.77, 11.9] & [8.25, 12.5] \\
$g_{\rho N}$ & 12.2 & [11.5, 13.1] & [10.8, 15.1] & [11.3, 13.6] & [10.5, 16.4] & [9.88, 19.7] \\
$b_\sigma/10^{-3}$ & 4.44 & [3.47, 5.76] & [2.54, 8.89] & [3.11, 6.60] & [2.33, 10.5] & [1.91, 14.4] \\
$c_\sigma/10^{-3}$ & -4.74 & [-6.47, -3.15] & [-10.0, 0.151] & [-7.51, -2.24] & [-11.9, 2.28] & [-16.7, 12.5] \\
$\Lambda_V$ & 0.176 & [0.145, 0.214] & [0.112, 0.293] & [0.132, 0.235] & [0.102, 0.337] & [0.0795, 0.440] \\
$g_{\sigma^* \Lambda}$ & 6.95 & [5.74, 7.82] & [2.89, 8.57] & [5.07, 8.11] & [1.37, 8.77] & [0.124, 9.24] \\
$g_{\phi \Lambda}$ & 3.62 & [1.70, 5.97] & [0.311, 9.24] & [1.04, 7.22] & [0.137, 10.1] & [0.00983, 11.4] \\
$g_{\sigma R}$ & $\times$ & [$\times$, 0.979] & [$\times$, 2.45] & [$\times$, 1.30] & [$\times$, 3.90] & [$\times$, 5.41] \\
$g_{\omega R}$ & $\times$ & [$\times$, 0.773] & [$\times$, 3.94] & [$\times$, 1.09] & [$\times$, 9.73] & [$\times$, 11.3] \\
$g_{\rho R}$ & $\times$ & [$\times$, 0.922] & [$\times$, 4.98] & [$\times$, 1.29] & [$\times$, 10.8] & [$\times$, 14.2] \\
$g_{\sigma Y}$ & $\times$ & [$\times$, 1.05] & [$\times$, 2.82] & [$\times$, 1.52] & [$\times$, 3.62] & [$\times$, 7.42] \\
$g_{\omega Y}$ & 3.72 & [3.14, 4.32] & [2.22, 5.63] & [2.85, 4.66] & [1.77, 6.38] & [0.135, 7.85] \\
$g_{\rho Y}$ & 4.41 & [3.66, 5.22] & [2.53, 7.00] & [3.29, 5.69] & [1.99, 7.89] & [0.154, 10.1] \\
$g_{\sigma^* Y^{'}}$ & 5.50 & [4.30, 6.52] & [1.17, 7.65] & [3.57, 6.94] & [0.372, 7.97] & [0.0108, 8.75] \\
$g_{\phi Y^{'}}$ & $\times$ & [$\times$, 1.63] & [$\times$, 3.95] & [$\times$, 2.23] & [$\times$, 5.35] & [$\times$, 8.84] \\
\bottomrule
\end{tabular}
\end{table*}

\clearpage
\bibliography{references}

\end{document}